\documentclass{article}
\usepackage{moreverb}
\usepackage{import}
\usepackage{mathtools,amssymb}
\usepackage[sorting=none]{biblatex} 
\let\cite=\supercite
\usepackage{caption}
\usepackage{subcaption}
\usepackage{multirow}
\usepackage{multirow}
\usepackage{bbm}
\usepackage{PRIMEarxiv}
\usepackage{placeins}
\usepackage{chngpage}

\usepackage[utf8]{inputenc} 
\usepackage[T1]{fontenc}    
\usepackage{hyperref}       
\usepackage{url}            
\usepackage{booktabs}       
\usepackage{amsfonts}       
\usepackage{nicefrac}       
\usepackage{microtype}      
\usepackage{lipsum}
\usepackage{fancyhdr}       
\usepackage{graphicx}       
\graphicspath{{media/}}     

	\title{The use of restricted mean survival time to estimate treatment effect under model misspecification -- a simulation study}

\author{
  Emily Alger\\
  Clinical Trial and Statistics Unit \\
  Institute of Cancer Research \\
  London, UK\\ 
  MRC Biostatistics Unit \\
  University of Cambridge\\
  Cambridge, UK\\
  \texttt{emily.alger@icr.ac.uk} \\
   \And
  David S.\ Robertson\\
  MRC Biostatistics Unit \\
  University of Cambridge\\
  Cambridge, UK\\
  \And
  Abigail J.\ Burdon\\
  MRC Biostatistics Unit \\
  University of Cambridge\\
  Cambridge, UK\\
}

\begin{document}
\maketitle

\begin{abstract}
The use of the non-parametric Restricted Mean Survival Time endpoint (RMST) has grown in popularity as trialists look to analyse time-to-event outcomes without the restrictions of the proportional hazards assumption. In this paper, we evaluate the power and type I error rate of the parametric and non-parametric RMST estimators when treatment effect is explained by multiple covariates, including an interaction term. Utilising the RMST estimator in this way allows the combined treatment effect to be summarised as a one-dimensional estimator, which is evaluated using a one-sided hypothesis Z-test. The estimators are either fully specified or misspecified, both in terms of unaccounted covariates or misspecified knot points (where trials exhibit crossing survival curves). A placebo-controlled trial of Gamma interferon is used as a motivating example to simulate associated survival times. 
    When correctly specified, the parametric RMST estimator has the greatest power, regardless of the time of analysis. The misspecified RMST estimator generally performs similarly when covariates mirror those of the fitted case study dataset. However, as the magnitude of the unaccounted covariate increases, the associated power of the estimator decreases. In all cases, the non-parametric RMST estimator has the lowest power, and power remains very reliant on the time of analysis (with a later analysis time correlated with greater power). 
\end{abstract}


	\maketitle

\section{Introduction}\label{sec1}
For many disease types, including cancers, randomised controlled trials (RCTs), which consider overall survival (OS) as the primary endpoint, are widely accepted as the gold standard for testing a new experimental treatment against control in Phase III trials. Originally, the log-rank statistic was utilised to  estimate the hazard ratio for time-to-event (TTE) outcomes due to its interpretability and distributional results. Such Phase III studies involved a large number of patients. As trials evolved, with a new emphasis on minimising patient risk and maximising patient benefit, Cox proportional hazards models became the new standard for the analysis of TTE outcomes  analyse. This semi-parametric method allows for greater flexibility and inclusion of covariates.\cite{cox1972regression} Generally, covariate adjustment leads to higher power and fewer required patients within Phase III trials.\cite{rosenbaum2002covariance} 

In recent years, there has been increasing concern regarding the proportional hazards assumption. An imbalance in patient covariates after randomisation means that often the hazard rate is not  constant over time.\cite{stensrud2019limitations} The Food and Drug Administration (FDA) have published guidelines concerning inclusion of baseline covariates in TTE studies.\cite{us2019guidance} To overcome this, many trialists have adopted the Restricted Mean Survival Time (RMST) as an alternative measure to the hazard ratio. As introduced by Royston et al. (2011),~\cite{ROYST2011} the RMST is evaluated as the area under the survival curve up to time $t^*$ and the difference in RMST between treatment arms is used as an estimator of the treatment effect.

The non-parametric RMST framework is most commonly used to mitigate the challenges relating to model misspecification since it requires no assumptions regarding the shapes of the survival curves for each treatment arm. However, our interest in RMST is motivated by trials that include multiple sources of information about the treatment effect. This occurs when multiple endpoints are influenced by treatment\cite{bauer1991multiple} and also for causal inference problems where the outcome depends on treatment via multiple pathways.\cite{lipkovich2020causal} Within this paper, we consider the RMST as an efficient summary measure to combine a multivariate vector of parameter estimates to form a univariate test statistic. For this reason, we employ the parametric RMST to a model that includes an interaction term between treatment and an indicator of inheritance pattern. This represents a causal inference problem where a pattern of inheritance moderates the effects of treatment on the TTE outcome.

In this paper, we compare multiple parametric and non-parametric methods to calculate treatment estimators using RMST. We look to identify the method that maximises the power of the trial through the use of covariates, yet is robust to model misspecification. In this instance, we compare the non-parametric RMST estimator to the best case scenario where survival time does indeed follow the proportional hazard assumption -- providing an upper bound on possible observed power and type I error estimates. We extend simulation studies to consider the impact of unaccounted covariates and also misspecified knot-points for a trial where survival curves cross over. We give consideration to the choice of the truncation time $t^*$ and evaluate identifiability issues for the different methods. 
In Section \ref{sec2}, we present the theory required to estimate the RMST estimand using a non-parametric and parametric method, and the formulation of the associated hypothesis test. In Section \ref{sec3}, we present the case study used to evaluate the success of RMST estimators. Section \ref{sec4}, investigates the impact of model misspecification on the RMST estimator whilst Section \ref{sec5} similarly investigates the misspecification of knot points when treatments are simulated so that their survival curves cross. A final discussion of results and extensions to this work is presented in Section \ref{sec6}.

\section{Statistical Theory}\label{sec2}
\subsection{Restricted Mean Survival Time}
\label{sec:RMST_theory}
As presented by Royston and Parmar \cite{ROYST2011}, the Restricted Mean Survival Time (RMST) is defined as the area below a survival curve $S(t)$ and can be used as an outcome measure for time-to-event clinical trials. For some time point $t^*$, the RMST $\mu(t^*)$ is defined as,
$$\mu(t^*)=\int_0^{t^*} S(t) \; dt.$$
Intuitively,  $\mu(t^*)$ represents expected life expectancy until time $t^*$ \cite{ROYST2011}. 
Within this paper we consider RMST as an estimator to measure treatment effect. To determine whether a treatment has a statistically significant effect, we consider the difference in life expectancy for two distinct treatments by evaluating the difference and variance of the  RMST estimator until time $t^*$. For two treatment arms with associated survival functions $S_1 \text{ and } S_2$, the RMST difference is defined as,
$$\Delta(t^*) = (\mu_2-\mu_1) (t^*) = \int_0^{t^*} S_2(t)-S_1(t) \; dt.$$

\subsection{Non-parametric RMST}

Non-parametric RMST  is calculated by evaluating the area under the non-parametric Kaplan-Meier estimate of survival $S(t)$. 

The Kaplan-Meier estimate of survival at time $t$ is defined as,
$$S(t) = \prod_{t_i<t}\left(1-\frac{d(t_i)}{n(t_i)}\right),$$
where $d(t_i)$ is the number of patients who experience an event and $n(t_i)$ is the total number of patients in the study at time $t_i$. 

For $t_i \in \{t_1,...,t_N\}$ such that $t_N \leq t^* < t_{N+1}$, the non-parametric RMST at time $t^*$ is defined as,

$$\mu(t^*) = \left (\sum_{t_i \leq t^*} S(t_i)(t_i - t_{i-1}) \right ) + S(t_N) (t^* - t_N),$$

where $t_0=0$.
It is of note that calculating non-parametric RMST relies on no model assumptions. 

\subsection{Parametric RMST}\label{sec:param_rmst}
For Cox proportional hazard models, the hazard associated with survival is characterised by some baseline hazard function $\lambda_0(t)$ (which can vary across time) and a component describing the way covariates effect the hazard, $\exp(\boldsymbol{\beta}^{T}\textbf{x})$. The hazard for a Cox proportional hazard model is given by,
\begin{align*}
    h(t|\textbf{x}) &= \lambda_0(t)\exp(\boldsymbol{\beta}^{T}\textbf{x}). \\
\end{align*}

The cumulative hazard function $H(t|\textbf{x})$ describes the hazard accumulated from the beginning of the trial to time $t$ and is defined as,
\begin{align*}
    H(t|\textbf{x}) &= \int_0^t h(u|\textbf{x}) \; du = \int_0^t \lambda_0(u) \exp(\boldsymbol{\beta}^{T} \textbf{x}) \; du. 
\end{align*}

The survival function $S(t)$, describing the probability that a patient survives up until at least time $t$, has the following relationship with the cumulative hazard,
\begin{align*}
\mathbb{P}(T>t) = S(t|\textbf{x})  &= \exp(-H(t|\textbf{x})).
\end{align*}

In this paper, we model survival using a Cox-exponential hazard where $\lambda_0(t)$ is defined as $\lambda$.
In this case, the survival function $S(t)$ is defined as
$$S(t)  = \exp(-\lambda t \exp(\boldsymbol{\beta}^T \textbf{x})).$$

The fully specified RMST is calculated using all relevant covariates contained in $\textbf{x}$. In this case, for some parameter vector $\boldsymbol{\beta}$, the fully specified RMST is given by:
\begin{align*}
    \mu(t^*)&= \int_0^{t^*} S(t) \; dt =\int_0^{t^*} \exp(-\lambda t \exp(\boldsymbol{\beta}^T \textbf{x})) \; dt = \frac{1}{\lambda \exp(\boldsymbol{\beta}^T \textbf{x})}(1-\exp(-\lambda t^* \exp(\boldsymbol{\beta}^T \textbf{x}))).
\end{align*}

In this paper we will misspecify the proportional hazard model by omitting covariates which explain the hazard within a working model.   Misspecification of the proportional hazard regression model in this way may introduce bias into analysis and lead to a violation of the proportional hazard assumption\cite{STRUT1986}. Within this paper we look to evaluate the impact of model misspecification on the associated power and type I error of a misspecified RMST estimator.
The misspecified RMST is fit similarly to the calculation for the fully specified RMST but neglects one covariate.

\subsection{Evaluating RMST difference}\label{sec:z_value}
When fitting a model with a treatment variable and additional treatment interaction term, we have two parameters quantifying treatment effect. The RMST estimator has the potential to combine these two parameters into one treatment estimator which can be used in a hypothesis test. Henceforth in this paper, this combined treatment effect (explained both by the treatment and interaction coefficient) will be described as the marginal treatment effect.
To evaluate whether RMST can be used to determine treatment efficacy we consider a one-sided hypothesis test at the 2.5\% level. We perform a one-sided hypothesis test on the difference in treatment effect for three methods -- calculating RMST using a non-parametric approach and calculating RMST using a misspecifed or fully specified parametric approach. 

The true RMST difference at time $t^*$ is defined as, 
$$\Delta(t^*) = \mu_2(t^*) - \mu_1(t^*),$$
where $\mu_1(t^*)$ and $\mu_2(t^*)$ are the RMST estimands for the control and treatment arm respectively.

We shall test the following hypothesis:

\begin{align*}
    \text{H}_0: \Delta(t^*) =0,  \\
    \text{H}_1: \Delta(t^*) >0.
\end{align*}
In words, the null hypothesis, $\text{H}_0$, is that there is no difference in marginal treatment effect between the two treatment arms. The alternate hypothesis, $\text{H}_1$, is that the treatment is efficacious compared to a control.

We test this hypothesis by calculating treatment estimates and associated variances using an  RMST estimator in order to produce Z-statistics. These are then compared to standard normal quantiles.  
The Z-statistics calculated in this simulation study are defined as,
$$Z = \frac{\hat{\Delta}(t^*)}{\text{s.e.}(\hat{\Delta}(t^*))}.$$

Whereas $\Delta(t^*)$ defines the true RMST difference, $\hat{\Delta}(t^*)$ represents the estimated treatment effect of the RMST estimator and `s.e.' represents the estimated standard error of RMST estimator. A description of how $\hat{\Delta}(t^*)$ and $\text{s.e.}(\hat{\Delta}(t^*))$ are estimated within this paper is presented in Sections \ref{method:np} to \ref{method:delta}. 
For each parametric and non-parametric RMST estimator $\hat{\Delta}(t^*)$, we confirm that the Z-statistics associated with marginal treatment effect are normally distributed by inspecting the histogram and Q-Q plots. Associated histograms and Q-Q plots for $N=10^4$ simulations of Z-statistics under each RMST estimator are presented in Section \ref{secs2} of the supplementary materials.

\section{Case study}\label{sec3}
\subsection{Working Model}

To evaluate the RMST estimator, we consider the `cgd' dataset within the `survival' package \cite{THERN2022} in R\cite{R} as a case study. The `cgd' dataset reports results of a randomised controlled trial of Gamma interferon. It evaluates time to serious infection observed in a population of 128 patients with chronic granulomatous disease. As each patient may have multiple infections within the observation window, we only consider the time to first serious infection for each patient.  

We fit a Cox-exponential model to predict time to serious infection using the `eha' package \cite{BROST2012} in R. We fit the model with five covariates -- some scalar baseline hazard $\lambda$, and four explanatory binary variables: `treatment'(treatment or control) `inherit' (pattern of inheritance), `sex' (male or female), and a `treatment:inherit' interaction. 
The factor levels of each variable are presented in Table \ref{factor} and parameter values associated for the Cox-exponential model for the `cgd' dataset is presented in Table \ref{cgd}.

\begin{table}[!htb]
\centering
\begin{tabular}{l|l|l}
\textbf{Variable}                   & \textbf{Factor level}     & \textbf{Level value} \\ \hline
\multirow{2}{*}{$X_\text{Treatment}$} & Control          & 0           \\ \cline{2-3} 
                           & Gamma interferon & 1           \\ \hline
\multirow{2}{*}{$X_\text{Inherit}$}   & X-linked         & 0           \\ \cline{2-3} 
                           & Autosomal        & 1           \\ \hline
\multirow{2}{*}{$X_\text{Sex}$}       & Male             & 0           \\ \cline{2-3} 
                           & Female           & 1           \\ 
\end{tabular}
\caption{Variable factor levels for the `cgd' dataset. \label{factor}}
\end{table}

\subsection{Simulating clinical trials} \label{sim}
To investigate the performance of the RMST estimator with an interaction term, we conduct a simulation study to evaluate the power and type I error rate of each RMST estimator. 
To generate survival times $T$ which are representative of the working model, we use the following equation,\cite{BENDE2005}
$$T=\frac{-\log(U)}{\lambda \exp(\boldsymbol{\beta}^T\textbf{x})},$$
where $U \sim \text{Uniform}[0,1]$.

Patients are assigned to treatment, sex and inheritance pattern using independent Bernoulli(0.5) random variables. The parameter values $\boldsymbol{\hat{\beta}}$ used to generate survival times are presented in Table \ref{cgd}, and are fitted using the original dataset under a Cox-exponential model with the R package `eha'. \cite{BROST2012}   

\begin{table}[!htb]
\centering
\begin{tabular}{l|l|l}
\textbf{Variable}                                & \textbf{Coefficient value}      & \textbf{Explanation}        \\ \hline
$\hat{\lambda}: \text{ Base hazard}$          & 0.015777 & Base hazard        \\ \hline
$\hat{\beta}_1: \text{ Treatment}$            & -1.116749  & Received treatment \\ \hline
$\hat{\beta}_2: \text{ Inherit}$              & 0.094373 & Autosomal          \\ \hline
$\hat{\beta}_3: \text{ Sex}$                  & -0.402188 & Female             \\ \hline
$\hat{\beta}_{12}: \text{ Treatment:Inherit}$ & 0.475445  & Treatment:Female   \\ 
\end{tabular}
\caption{Variables and coefficient values used to predict time to serious infection.}
\label{cgd}
\end{table}

Random censoring is simulated using an exponential random variable with rate 0.001, ensuring that approximately 10\% of patients are randomly censored. The arrival window for patients to the study was 20 weeks and the study analysis occurs at 120 weeks so that maximum follow-up was 120 weeks. With these parameter values, overall censoring is approximately 40\%. We fix trial sample size at 100 such that under a simulation study of $N=10^5$ trials, fully-specified parametric power is approximately 90\%. 

\section{Investigating misspecified parametric survival models}\label{sec4}
\subsection{Estimation of non-parametric RMST difference}\label{method:np}
Within this paper, the non-parametric RMST is determined using the `rmst2' function in the `survRM2' package \cite{UNO2022} in R. The variance of the non-parametric RMST is calculated using the `survRM2'\cite{UNO2022} function which uses the Greenwood's plug-in estimator for asymptotic variance. 

\subsection{Estimation of fully specified parametric RMST difference}\label{method:fs}
To calculate the parametric RMST, we derive the area under a Cox-exponential survival curve similarly to Section \ref{sec:param_rmst}, given estimated parameter values $\boldsymbol{\hat{\theta}}$ and $\hat{\lambda}_1$,

\begin{equation}
    \begin{split}
        \hat{\mu}_\text{full}(t^*|\boldsymbol{\hat{\theta}}, \hat{\lambda}_1,\mathbf{X}) = \frac{(1-\exp(-\hat{\lambda}_1 t^* \exp(\hat{\theta}_1 X_\text{Treatment}+\hat{\theta}_2 X_\text{Inherit}+\hat{\theta}_3 X_\text{Sex}+\hat{\theta}_{12} X_\text{Treatment:Inherit})))}{\hat{\lambda}_1 \exp(\hat{\theta}_1 X_\text{Treatment}+\hat{\theta}_2 X_\text{Inherit}+\hat{\theta}_3 X_\text{Sex}+\hat{\theta}_{12} X_\text{Treatment:Inherit})}. \\  
    \end{split}
    \label{equ:full}
\end{equation}

\noindent The RMST difference estimator is defined as,

$$\hat{\Delta}_\text{full}(t^*|\boldsymbol{\hat{\theta}}, \hat{\lambda}_1) = \hat{\mu}_\text{full}(t^*|\hat{\boldsymbol{\theta}},\hat{\lambda}_1,X_\text{Treatment}=1)-\hat{\mu}_\text{full}(t^*|\boldsymbol{\hat{\theta}},\hat{\lambda}_1,X_\text{Treatment}=0).$$
The estimated RMST is calculated using parameter estimates fitted using maximum likelihood estimation as implemented by the `phreg' function in R package `eha' \cite{BROST2012}.
A full derivation of $\hat{\Delta}_\text{full}(t^*|\hat{\boldsymbol{\theta}}, \hat{\lambda}_1)$ under our simulation scenario is presented in the Supplementary materials, Section \ref{secs1}.

\subsection{Estimation of misspecified parametric RMST difference}\label{method:ms}
Under the working model used to simulate trials, survival is dependent on the `sex' covariate. To evaluate the effect of misspecification on the power of the parametric RMST estimator, we misspecify the estimator by neglecting the sex covariate during model fitting.

Given parameter estimates $\boldsymbol{\hat{\gamma}}$ and $\hat{\lambda}_2$, the misspecified parametric RMST is defined as,

\begin{equation}
    \begin{split}
        \hat{\mu}_\text{miss}(t^*|\boldsymbol{\hat{\gamma}}, \hat{\lambda}_2,\mathbf{X})) = \frac{(1-\exp(-\hat{\lambda}_2 t^* \exp(\hat{\gamma}_1 X_\text{Treatment}+\hat{\gamma}_2 X_\text{Inherit}+\hat{\gamma}_{12} X_\text{Treatment:Inherit})))}{\hat{\lambda}_2 \exp(\hat{\gamma}_1 X_\text{Treatment}+\hat{\gamma}_2 X_\text{Inherit}+\hat{\gamma}_{12} X_\text{Treatment:Inherit})}. \\
    \end{split}
\end{equation}
The estimator we evaluate in these simulations is the RMST difference, defined as,

$$\hat{\Delta}_\text{miss}(t^*|\boldsymbol{\hat{\gamma}} \hat{\lambda}_2) = \hat{\mu}_\text{miss}(t^*|\boldsymbol{\hat{\gamma}},\hat{\lambda}_2,X_\text{Treatment}=1)-\hat{\mu}_\text{miss}(t^*|\boldsymbol{\hat{\gamma}},\hat{\lambda}_2,X_\text{Treatment}=0).$$

\subsection{Delta Method}\label{method:delta}
 We utilise the delta method to approximate the variance of the fully specified and misspecified parametric RMST difference. 
 This method approximates the variance as,
 $$\text{Var}((\mu_2-\mu_1)(t^*)) \approx \nabla((\mu_2-\mu_1)(t^*))^{T} \; \frac{\Sigma}{n} \; \nabla((\mu_2-\mu_1)(t^*)).$$
 where $n$ is the number of covariates, $\Sigma$ is the variance-covariance matrix for these variables, and,
 
\begin{align}
    \nabla((\hat{\mu}_2-\hat{\mu}_1)(t^*)) = \left [ \frac{\partial((\hat{\mu}_2-\hat{\mu}_1)(t^*))}{\partial(\hat{\lambda})}, \frac{\partial((\hat{\mu}_2-\hat{\mu}_1)(t^*))}{\partial(\hat{\beta}_1)}, \hdots, \frac{\partial((\hat{\mu}_2-\hat{\mu}_1)(t^*))}{\partial(\hat{\beta}_n)} \right ]^T.
\end{align}

\subsection{Simulation study set-up}
\label{sec:simulation_noncross}
To evaluate the power of the RSMT difference as an estimator of treatment efficacy in the presence of a treatment interaction, we ran a simulation study. 
The working model that we simulated clinical trials from is a Cox-exponential model, with fitted coefficient values as presented in Table \ref{cgd} and hazard,
$$h(t|\textbf{x})=\hat{\lambda} \exp(\hat{\beta}_1 X_\text{Treatment}+\hat{\beta}_2 X_\text{Inherit}+\hat{\beta}_3 X_\text{Sex}+\hat{\beta}_{12} X_\text{Treatment:Inherit}).$$

\begin{figure}[ht!]
\centering\includegraphics[width=\columnwidth]{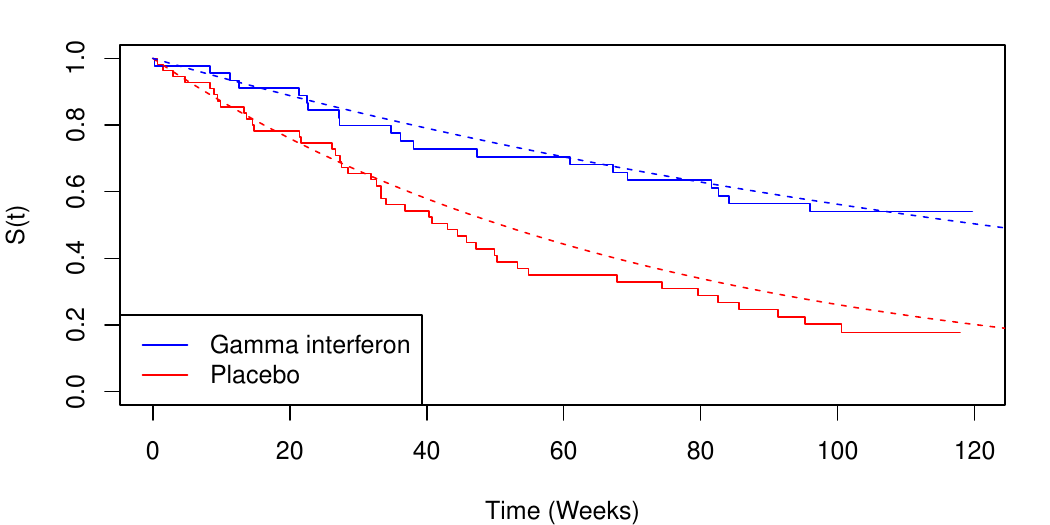}
\caption{A Kaplan-Meier plot presenting the survival of 100 patients on control or Gamma interferon treatment. Overlayed dashed lines show the parametric equation used to simulate patient survival times.}
\label{fig:example_trial}
\end{figure}

An example trial simulated with these parameter values is presented in Figure \ref{fig:example_trial}.

We generate clinical trial data using the method presented in Section \ref{sim}. Using a hypothesis test described in Section \ref{sec:z_value}, we evaluate whether we would reject the null hypothesis that both treatment arms had the same efficacy effect using three methods of estimating RMST: using a non-parametric, misspecified or fully specified parametric approach. 

We repeat this process and generate $N=10^5$ clinical trials.
To evaluate power, we simulate trials with a `treatment' and `treatment:inherit' interaction effect as presented in Table \ref{cgd}. The power is calculated as the proportion of all trials where we correctly reject the null hypothesis when there is a treatment effect.  

To evaluate the type I error rate, we simulate trials where the `treatment' parameter and `treatment:inherit' interaction are forced to be zero, but all other parameters remain the same as in Table \ref{cgd}. The type I error rate was calculated as the proportion of all trials where we incorrectly reject the null hypothesis.

We assess the power and type I error rates as $t^*$ varies from 1 to 150 weeks with $\beta_3$ fixed at -0.40. We also evaluate these rates as $\beta_3$ varies from -2 to 2 and $t^*$ is fixed at 100 weeks. Note that for a sample size of 100 patients, when $\beta_3$ is negative it is not always the case that every simulated clinical trial realisation will have all combinations of treatment, pattern of inheritance, and sex. In these cases, we cannot calculate maximum likelihood estimations of coefficients and the specific simulated trial realisation is not used to estimate power and type I error rates. When $\beta_3$ is at its most negative value of -2, 249 trials cannot be used to estimate power or type I error, representing 0.2\% of all trials simulated for power calculations. The number of trials which have to be discarded for power calculations decreases steadily as $\beta_3$ becomes more positive. When $\beta_3 \geq 0.22$, no trials need to be discarded. 

\subsection{Simulation results}
\label{subsec:measure_comparison}

\begin{figure}[htb]
     \centering
     \begin{subfigure}[t]{0.49\textwidth}
         \centering
\includegraphics[width=\textwidth]{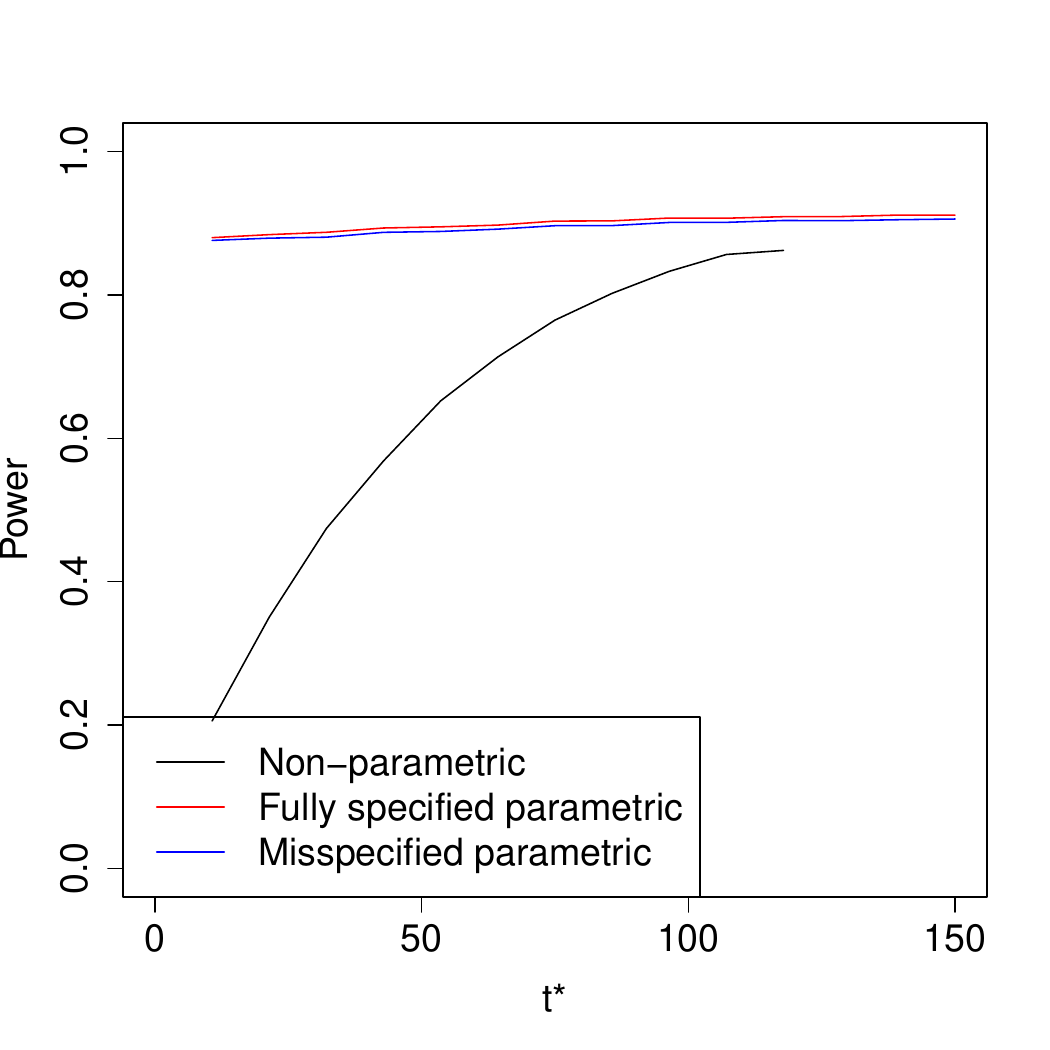}
         \label{fig:t_star}
         \caption{Power for varying $t^*$ with $\beta_3=-0.402$.}
     \end{subfigure}
     \hfill
     \begin{subfigure}[t]{0.49\textwidth}
         \centering
\includegraphics[width=\textwidth]{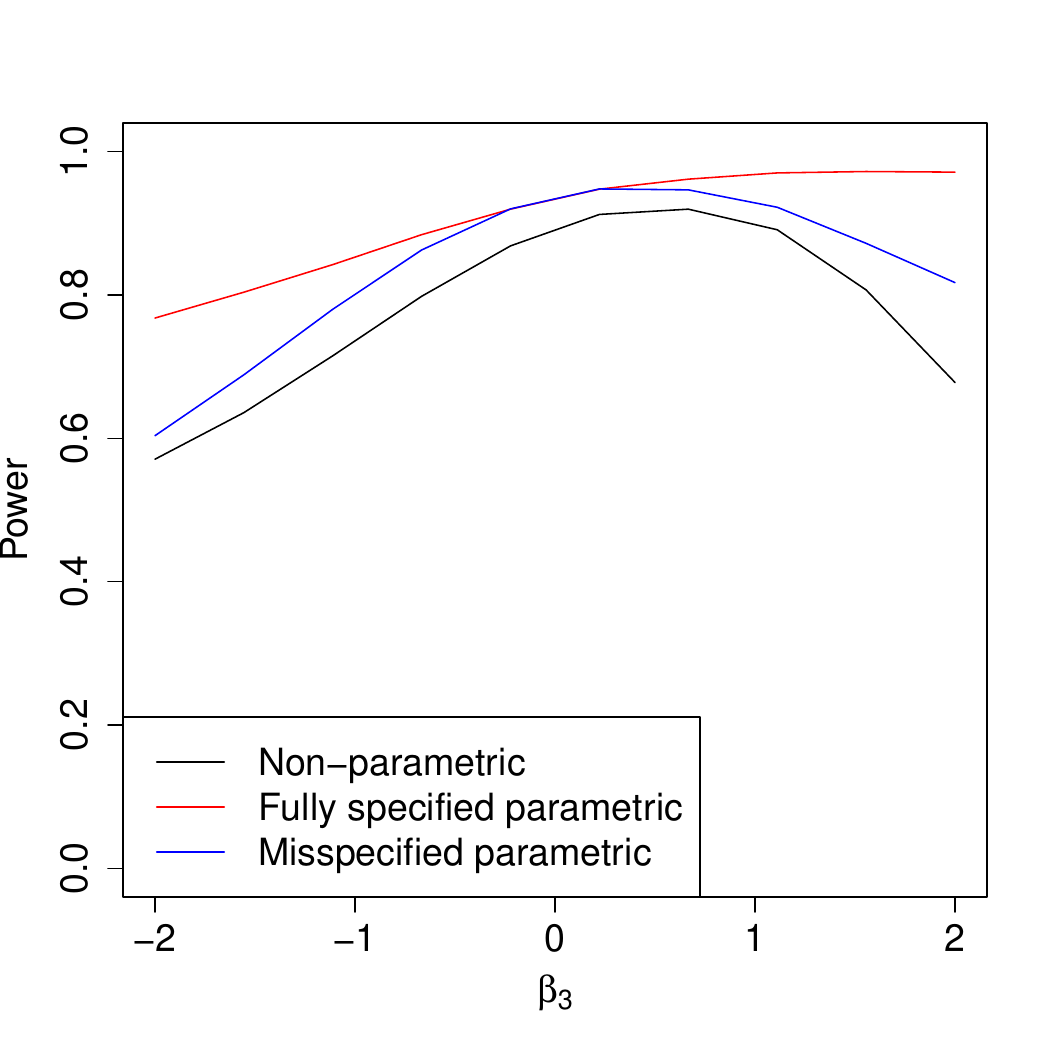}
         \label{fig:cov_3}
         \caption{Power for varying  $\beta_3$ with $t^*=100$.}
     \end{subfigure}
        \caption{Simulation study results showing achieved power when data is fit to a Cox proportional hazards model.}
        \label{fig:power}
\end{figure}

Non-parametric RMST difference can only be calculated for a pre-specified $t^*$ if there is at least one observable patient on each arm of treatment at that time point or beyond. Therefore non-parametric RMST difference can only be used as an estimator of treatment efficacy until, at the very latest, the end of the trial (120 weeks). Trials where the last observable patient experienced an event before $t^*$ are excluded from power and type I error rate calculations. 
As $t^*$ becomes larger, in order to evaluate the RMST estimator, the last observable patient needs to remain on the trial for longer. Therefore, as $t^*$ increases, more trials are excluded from the non-parametric power calculations. 42 trials are excluded from power calculations at  107 weeks.  By 118 weeks, 60,549 of trials (60.5\%) cannot be used as part of power calculations. A similar pattern emerges when power is evaluated as $\beta_3$ varies. In this case, as $\beta_3$ becomes more positive, and more events are observed, patients are less likely to remain on the trial until 100 weeks. When $\beta_3$ is set at 0.2, 57 trials (0.057\%) cannot be used for power calculations. This increases steadily until $\beta_3$ is set to 2, where 1,223 trials (1.2\%) cannot be used to evaluate power. 
The power and type I error rate as $t^*$ and $\beta_3$ varies is presented in Figures \ref{fig:power} and \ref{fig:type1}. 

\begin{figure}[ht!]
     \centering
     \begin{subfigure}[t]{0.49\textwidth}
         \centering
         \includegraphics[width=\textwidth]{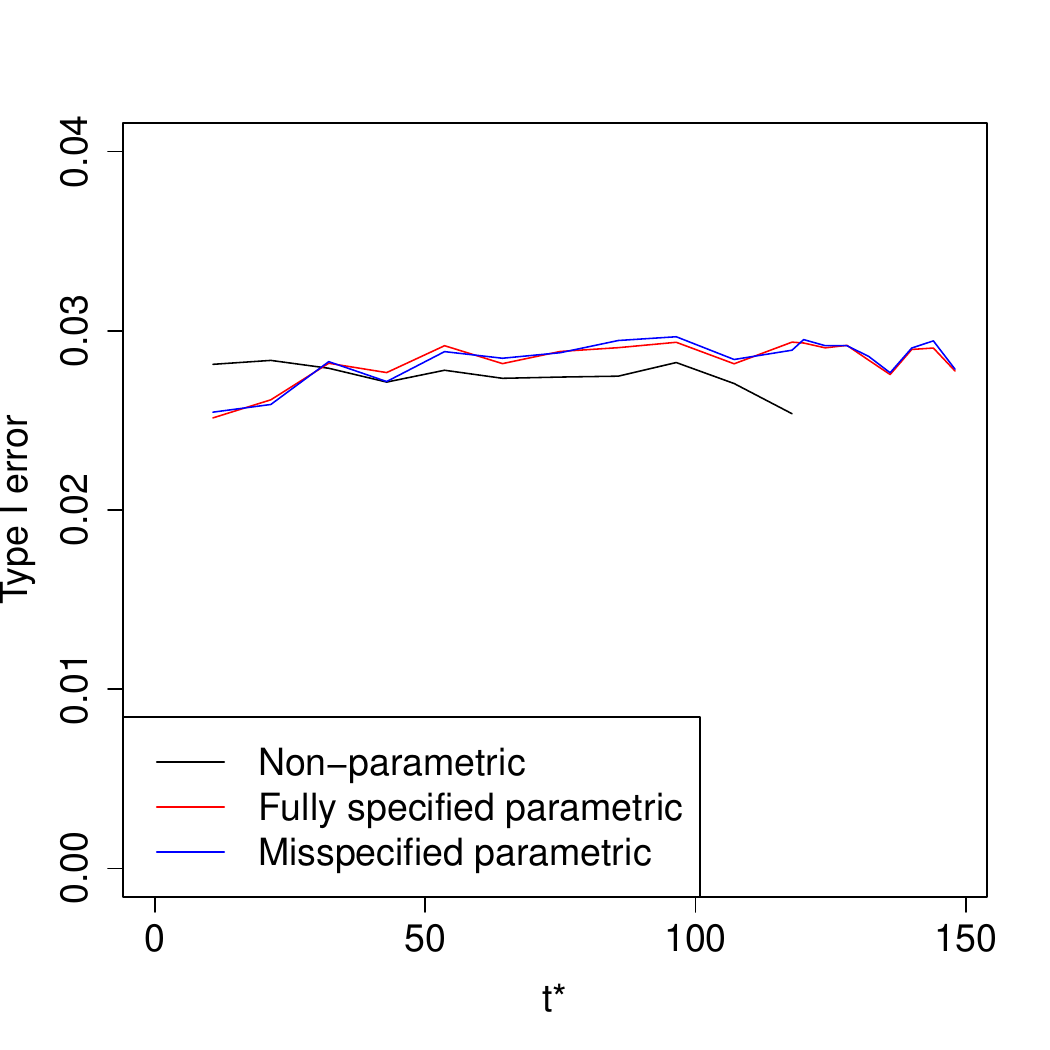}
         \label{fig:t_starerror1}
         \caption{Type I error for varying $t^*$ $\beta_3=-0.402$.}
     \end{subfigure}
     \hfill
     \begin{subfigure}[t]{0.49\textwidth}
         \centering
         \includegraphics[width=\textwidth]{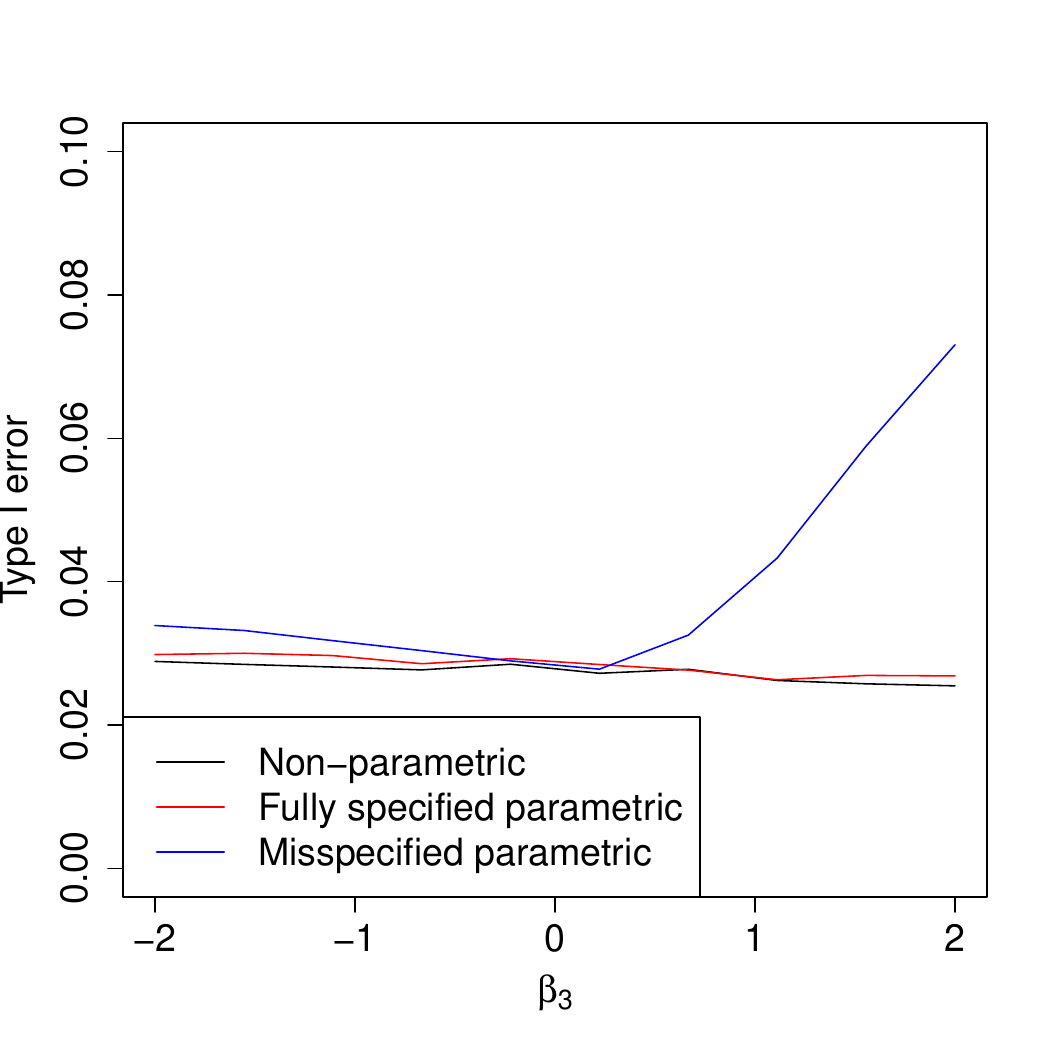}
         \label{fig:type1}
         \caption{Type I error for varying  $\beta_3$ with $t^*=100$.}
         \label{fig:cov_3error1}
     \end{subfigure}
        \caption{Simulation study results showing type I error  when data is fit to a Cox proportional hazards model.}
         \label{fig:type1}
\end{figure}

\subsubsection{Non-parametric estimator}
For the trials where RMST difference could be calculated, as $t^*$ increases the difference in the area under the curve for the treatment and control increases, thus increasing the magnitude of the marginal treatment effect.  The associated standard error increases but at a lesser rate, thus increasing the magnitude of the Z-statistic. This increases the probability that the hypothesis test will reject the null hypothesis and thus increases the power. 
When $t^*$ is fixed, the power of RMST estimator decreases as the magnitude of $\beta_3$ increases. As the magnitude of $\beta_3$ increases, the treatment survival curves become closer together and the difference in RMST difference decreases. The standard error remains stable as the value of $\beta_3$ varies. 
Type I error rates for the non-parametric estimator remains relatively stable with varying $t^*$ and $\beta_3$ values.

\subsubsection{Misspecified parametric estimator}
The power of the misspecified parametric RMST estimator remains stable across all values of $t^*$. The difference in the area under the survival curve and the standard error of the associated statistic increases proportionately, keeping the value of the Z-statistic approximately constant.
When we fix $t^*$ and increase the magnitude of the sex coefficient, the power decreases similarly to  the non-parametric RMST estimator. 
For the misspecified model, the type I error rate inflates above the expected 2.5\% to 7.3\% as $\beta_3$ becomes more positive. As $\beta_3$ becomes more positive, the standard error of the estimator deflates as more events are observed. This increases the magnitude of some Z-statistics, thus prompting more rejections of the null hypothesis and inflating the type I error rate.

\subsubsection{Fully specified parametric estimator }

Similarly to the misspecified parametric RMST estimator, the power of the fully specified parametric RMST estimator remains constant as $t^*$ increases. In particular, considering Equation \ref{equ:full}, as $t^*$ increases so too does the magnitude of the numerator of the parametric RMST estimator. 
The power of the fully specified parametric RMST difference also increases when we fix $t^*$ and increase $\beta_3$. Whilst the size of the marginal treatment effect decreases as the magnitude of $\beta_3$ increases, the standard error of the treatment effect decreases as $\beta_3$ becomes more positive. As $\beta_3$ becomes more positive, the hazard of incidence increases and so we have more incidence events at $t^*$. The greater number of incidence events provides more information and hence reduces the variance and increases the magnitude of the respective Z-statistic. Therefore the increase in power is driven by an increase in the number of observed events.
The Type I error for the non-parametric estimator remains relatively stable to varying $t^*$ and $\beta_3$ values.

\subsubsection{Comparison of results}

Unlike the parametrically defined RMST estimator, the non-parametric RMST estimator cannot extrapolate survival beyond the last observed failure time. Additionally, the non-parametric RMST power is sensitive to the value of $t^*$, slowly converging to the power of the parametric RMST estimator as $t^*$ tends toward the end of the trial. Therefore, both parametric RMST estimator's powers outperform that of the non-parametric RMST. For relatively small $\beta_3$, the fully specified and misspecified parametric RMST have similar power which remains relatively constant as $t^*$ varies over time. 

Whilst the power of the non-parametric RMST remains lower than that of the misspecified parametric RMST, they exhibit very similar behaviour when $t^*$ is fixed and we vary the size of $\beta_3$. In this setting, the fully-specified parametric measure outperforms both other estimators and power in fact increases as $\beta_3$ becomes more positive as more incident events occur. 

The type I error rate of the three methods remains stable as $t^*$ varies, with all RMST estimators having an empirical type I error rate of 2.8\%. There is additional randomness in terms of the type I error rate as $t^*$ extends beyond the end of the trial at 120 weeks. This variability may be explained by the extrapolation of survival beyond the observed survival window. 

As the value of $\beta_3$ varies for $t^*$ fixed at 100, the type I error rate associated with the non-parametric and fully specified RMST remains stable. The empirical type I error rate for the non-parametric and fully specified RMST estimator is 2.7\% and 2.8\%, respectively. Conversely, for the misspecified RMST, the type I error rate increases as the value of $\beta_3$ becomes more positive, with a largest type I error rate of 7.3\% when the value of $\beta_3$ reached 2.0. As $\beta_3$ becomes more positive, the variance of the estimator deflates. This may inflate the magnitude of some Z-statistics, thus giving rise to more rejections of the null hypothesis and inflating the type I error rate.

It is of note that all type I errors rates in these simulations are inflated above the expected 2.5\% error rate. These results reflect published findings \cite{DORMU2023}$^,$\cite{HORIG2020} that the type I error rate can be inflated due to small trial sample size (100 patients). For small sample sizes, the asymptotic assumption that the distribution of Z-statistics are normally distributed does not exactly hold. This increases the observed type I error rate by around 0.2\% in absolute terms. Further simulations (not shown) were undertaken to confirm this finding. At a sample size of 250 patients, all type I error rates are within the expected range according to a binomial distribution with $N=10^5$ trial simulations.

Sections \ref{secs3} and \ref{secs4} of the Supplementary materials contains graphical and tabulated mean summaries of Z-statistics, treatment effects, and associated standard error for power and type I error calculations as $t^*$ and $\beta_3$ varied.

\section{Investigating crossing survival curves}\label{sec5}
\subsection{Comparison using trial simulation}

In contrast to the usual Cox proportional hazards model, which estimates the survival distribution on each treatment arm at a single point in time, the RMST test statistic takes into account the complete shape of the survival function up to time \(t^*\). This is particularly desirable when it is known or assumed that the proportional hazards assumption does not hold. One such example is when the survival curves for different treatment arms cross over midway through the trial\cite{li2015statistical}. Clinical trial data displaying crossing survival curves is often indicative of a treatment that has a short-term benefit but no long-term benefit. For example, a surgical intervention may improve survival prospects for 6 weeks, yet over the course of 5 years, the non-surgical control treatment may prove to be more effective. We shall investigate the difference between the parametric and non-parametric RMST estimators when the true survival distributions cross over. We shall discuss the additional modelling assumptions that are required for such a trial.

In this scenario, similarly to Section~\ref{sec:simulation_noncross}, we consider the explanatory variables `treatment', `inherit', `sex' and also the interaction `treatment:inherit' status. Let \(\mathbf{X}=(X_\text{Treatment},X_\text{Inherit},X_\text{Sex},X_\text{Treatment:Inherit})^T\) be a \(4\times 1\) vector of covariates for a particular patient. At  a pre-specified time point $t_1$, the control treatment becomes more effective than the experimental treatment method. This time point \(t_1\) is also referred to as a `knot-point', as in the piecewise exponential model for survival data\cite{friedman1982piecewise}. In order to consider this, we allow each coefficient to take two distinct values, before and after \(t_1.\) The coefficient vectors, \(\boldsymbol{\beta}_a=(\beta_{1,a},\beta_{2,a},\beta_{3,a},\beta_{12,a})^T\) and \(\boldsymbol{\beta}_b=(\beta_{1,b},\beta_{2,b},\beta_{3,b},\beta_{12,b})^T, \) are each of the same dimension as \(\mathbf{X}\).  The hazard function, cumulative hazard function, and survival functions for crossing survival curves are respectively,
\begin{equation}
\label{eq:haz_cross}
\begin{split}
h_\text{cross}(t|\mathbf{X})&=
\lambda_a\exp(\boldsymbol{\beta}_a^T \mathbf{X})\mathbbm{1}\{t<t_1\} + \lambda_b\exp(\boldsymbol{\beta}_b^T \mathbf{X})\mathbbm{1}(1-\{t<t_1\}),
\end{split}
\end{equation}

\begin{equation}
\begin{split}
  H_{cross}(t|\mathbf{X}) &= 
\begin{dcases}
   \lambda_a t \exp(\boldsymbol{\beta}_a^T \mathbf{X}) & \text{if } t < t_1 \\
   \lambda_a t_1 \exp(\boldsymbol{\beta}_a^T \mathbf{X}) + \lambda_b (t-t_1) \exp(\boldsymbol{\beta}_b^T \mathbf{X}) & \text{otherwise}
\end{dcases}, \\
\notag
\smallskip
S_{cross}(t|\mathbf{X}) &= 
\begin{dcases}
   \exp\{-\lambda_a t \exp(\boldsymbol{\beta}_a^T \mathbf{X})\} & \text{if } t < t_1 \\
   \exp\{-\lambda_a t_1 \exp(\boldsymbol{\beta}_a^T \mathbf{X}) - \lambda_b (t-t_1) \exp(\boldsymbol{\beta}_b^T \mathbf{X})\} & \text{otherwise}
\end{dcases}.
\end{split}
\end{equation}
Following Section~\ref{sec:RMST_theory} and integrating the cumulative survival function up to \(t^*\), the RMST for the crossing curves is given by,
\begin{equation}
\label{equ:full_cc}
\mu_\text{cross}(t^*|\lambda_a,\lambda_b,\boldsymbol{\beta}_a,\boldsymbol{\beta}_b,\mathbf{X}) = 
\begin{dcases}
   \frac{(1-\exp\{-\lambda_a t^* \exp(\boldsymbol{\beta}_a^T \mathbf{X})\})}{\lambda_a\exp(\boldsymbol{\beta}_a^T \mathbf{X})} & \text{if } t^* < t_1 \\
  \frac{(1-\exp\{-\lambda_b t^* \exp(\boldsymbol{\beta}_b^T \mathbf{X})\})\exp\{\lambda_b t_1 \exp(\boldsymbol{\beta}_b^T \mathbf{X}-\lambda_a t_1 \exp(\boldsymbol{\beta}_a^T \mathbf{X}))}{\lambda_b\exp(\boldsymbol{\beta}_b^T \mathbf{X})} & \text{otherwise}.
\end{dcases}
\end{equation}

The fully specified, parametric crossing curves  estimator is the difference in RMST between treatment arms given by
\begin{equation}
    \label{equation:endpoint_cross}
    \Delta_\text{cross}(t^*)=\mu_\text{cross}(t^*|\hat{\lambda}_a,\hat{\lambda}_b,\hat{\boldsymbol{\beta}}_a,\hat{\boldsymbol{\beta}}_b,X_\text{Treatment}=1)-\mu_\text{cross}(t^*|\hat{\lambda}_a,\hat{\lambda}_b,\hat{\boldsymbol{\beta}}_a,\hat{\boldsymbol{\beta}}_b,X_\text{Treatment}=0),
\end{equation}
where $\hat{\lambda}_a,\hat{\lambda}_b,\hat{\boldsymbol{\beta}}_a,\hat{\boldsymbol{\beta}}_b$ are the maximum likelihood estimates for the parameters $\lambda_a,\lambda_b,\boldsymbol{\beta}_a,\boldsymbol{\beta}_b$ respectively. The estimates can be obtained using the `pch' package in R~\cite{frumento2021package}.
Parametric RMST is estimated similarly to Sections \ref{method:fs} and \ref{method:ms} when treatment effect was modified part way through the trial. 

\begin{figure}[ht!]
\centering\includegraphics[width=0.9\textwidth]{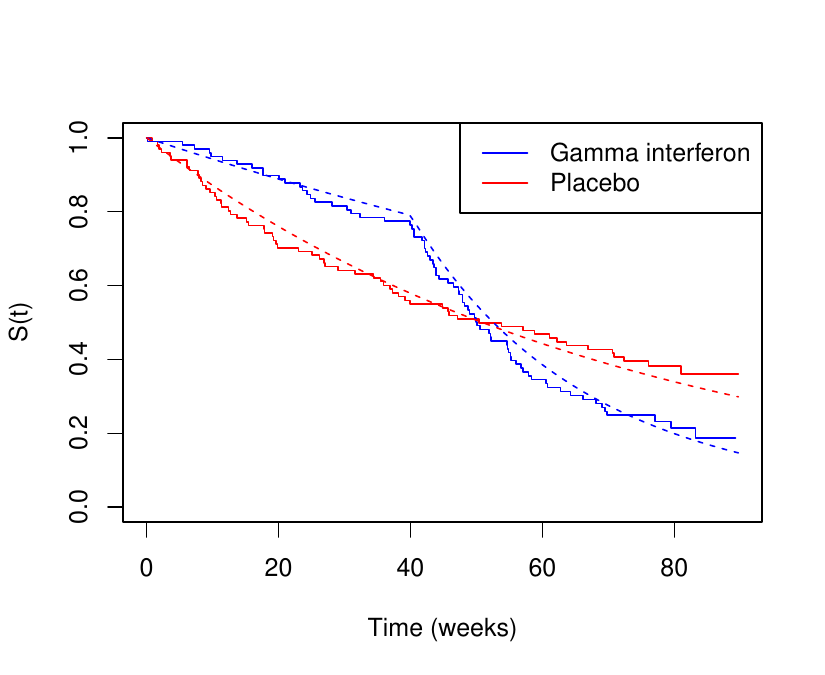}
\caption{A Kaplan-Meier plot showing the survival of 130 patients receiving either Gamma interferon or control treatment for survival distribution with crossing survival curves. Dashed lines show the parametric value of the survival function for the parameter values given in Table~\ref{tbl:cgd_cross}.}
\label{fig:cross_KM}
\end{figure}

In what follows, we shall simulate data in accordance with a Cox piecewise exponential model with a single knot point, Equation~\ref{eq:haz_cross}, and the parameters are informed by the `cgd' dataset. The true knot point of the piecewise model is fixed throughout as \(t_1=40\) and the treatment effects before and after \(t_1\), given by \(\beta_{1,a},\beta_{1,b}\), are altered and exaggerated to ensure that the survival curves cross. Table~\ref{tbl:cgd_cross} shows the values of the parameters used in the simulation studies. Figure \ref{fig:cross_KM} shows the Kaplan-Meier survival curves for a simulated dataset of \(n=130\) patients under this model. The dashed lines represent the true survival function \(S(t)\) averaged over covariate distributions for each of the treatments: Gamma interferon and control. In this example, the experimental treatment is effective in the short-term yet less effective than control in the long term.  We fix trial sample size at 130 such that under a simulation study of $N=10^4$ trials, fully-specified parametric power is approximately
90\%.

\begin{table}[ht!]
\centering
\begin{tabular}{l|l|l}
\textbf{Variable}          & \textbf{Coefficient value before time} \(t_1=40\)                 & \textbf{Coefficient value after time} \(t_1=40\)\\ \hline
Base Hazard       & \(\lambda_a=0.0158\)  & \(\lambda_b=0.0158\)        \\ \hline
Treatment         & \(\beta_{1,a}=-1.117\) & \(\beta_{1,b}=0.750\)           \\ \hline
Inherit           & \(\beta_{2,a}=0.094\) &       \(\beta_{2,b}=0.094\)     \\ \hline
Sex               & \(\beta_{3,a}=-0.402\)  &      \(\beta_{3,b}=-0.402\)       \\ \hline
Treatment*Inherit & \(\beta_{12,a}=0.475\) & \(\beta_{12,b}=0.475\)            \\ 
\end{tabular}
\caption{Values of coefficients used in the simulation study investigating crossing survival curves with a trial sample size of 130.}
\label{tbl:cgd_cross}
\end{table}

The RMST estimator is defined as the difference between areas under the survival curves of each treatment arm up to time \(t^*\). In Figure~\ref{fig:cross_KM}, although the knot point occurs at 40 weeks, there is a delay before the curves cross at roughly 55 weeks. Hence, as \(t^*\) increases, the RMST estimator is positive and increasing up to \(t^*=55\). The area between survival curves reaches net zero at roughly 90 weeks and hence as \(t^*\) increases, the RMST estimator is positive and decreasing between \(t^*=55\) and \(t^*=90\), and is negative and decreasing beyond \(t^* = 90.\) Therefore, we consider how the choice of \(t^*\) affects power for each of the parametric and non-parametric methods for the crossing survival curves model. We also note that \(t_1\) is fixed and defined prior to commencement of the trial. Methods are available which allow data driven estimation of \(t_1\)\cite{pons2002estimation}, however these methods are generally most appropriate for exploratory purposes and we have chosen to fix \(t_1\) to ensure that parameter estimation is robust and type I error rates are protected. For this simulation study, we shall investigate the robustness of the parametric and non-parametric RMST estimators under model misspecification through the value of the knot-point. That is, data is simulated using \(t_1=40\) but estimation of the parametric RMST estimator will use a range of \(\tilde{t}_1\) values in place of \(t_1\) in Equation~\ref{equ:full_cc}.

\subsection{Simulation results}

\begin{figure*}[!htb]
     \centering
     \begin{subfigure}[t]{0.49\textwidth}
         \centering
         \includegraphics[width=\textwidth]{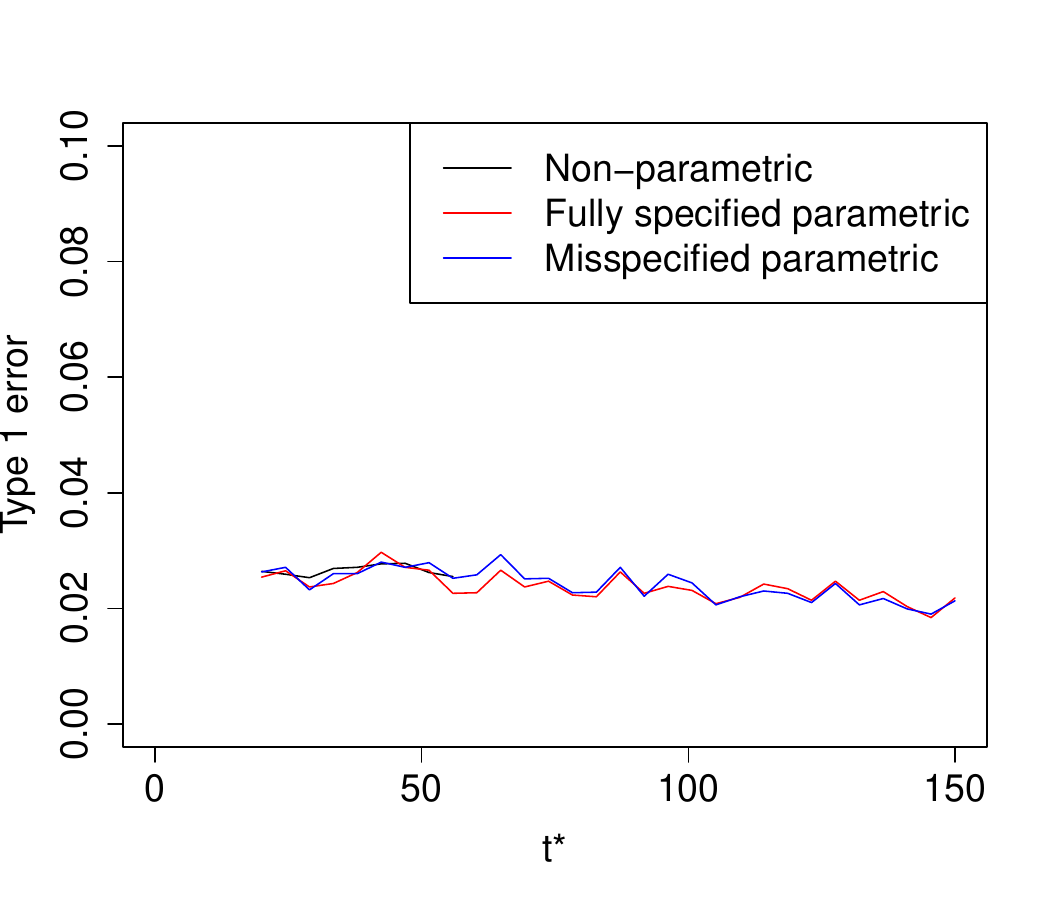}
         \caption{Varying $t^*$ with fixed true knot-point          $t_1=40$ and misspecified knot-point $\tilde{t}_1=50$.}
         \label{fig:type1_crossing_tstar}
     \end{subfigure}
     \hfill
     \begin{subfigure}[t]{0.49\textwidth}
         \centering
         \includegraphics[width=\textwidth]{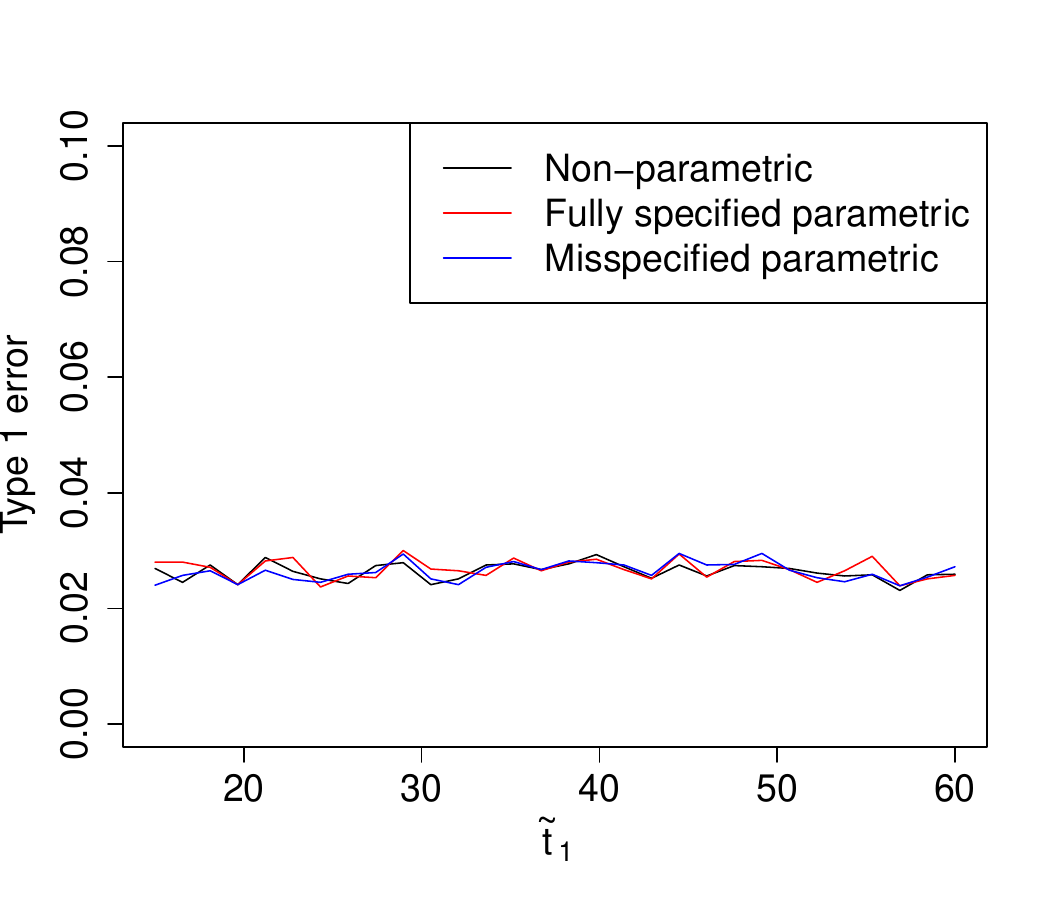}
         \caption{Varying misspecified knot-point $\tilde{t}_1$ with fixed $t^*=40$ and true knot-point $t_1=40$.}
         \label{fig:type1_crossing_t1}
     \end{subfigure}
        \caption{Simulation study results showing type I error rates when data is fit to a Cox proportional hazards model with crossing survival curves.}
\end{figure*}

Figure~\ref{fig:type1_crossing_tstar} summarises the effect of varying \(t^*\) on the type I error rates when a crossing survival curves model is assumed. For each point on the graph, \(N=10^4\) simulations were performed under the parameter values given in Table~\ref{tbl:cgd_cross} and with knot-point \(t_1=40\) used for simulations. For the misspecified model, the knot-point is assumed to be \(\tilde{t}_1\) at the model fitting stage. The results show that the type I error rate is protected at 2.5\% for all methods.  This is as expected since under \(H_0\), the survival curves are identical and do not cross so this feature is not affected by changes in \(t^*\). Further, these results are complimentary to the results of Section~\ref{subsec:measure_comparison}, however we note that there is more noise in Figure~\ref{fig:type1_crossing_tstar} than in Figure~\ref{fig:type1}. This is explained as we must fit twice the number of parameters because of the piecewise structure of the fitted model. Figure~\ref{fig:type1_crossing_t1} shows the effect of model misspecification on type I error rates. We find that the assumed value of the knot-point at the model fitting stage does not affect type I error. This is because under \(H_0\), the curves do not cross which mitigates the need to specify the knot point.

\begin{figure*}[!htb]
     \centering
     \begin{subfigure}[t]{0.49\textwidth}
         \centering
         \includegraphics[width=\textwidth]{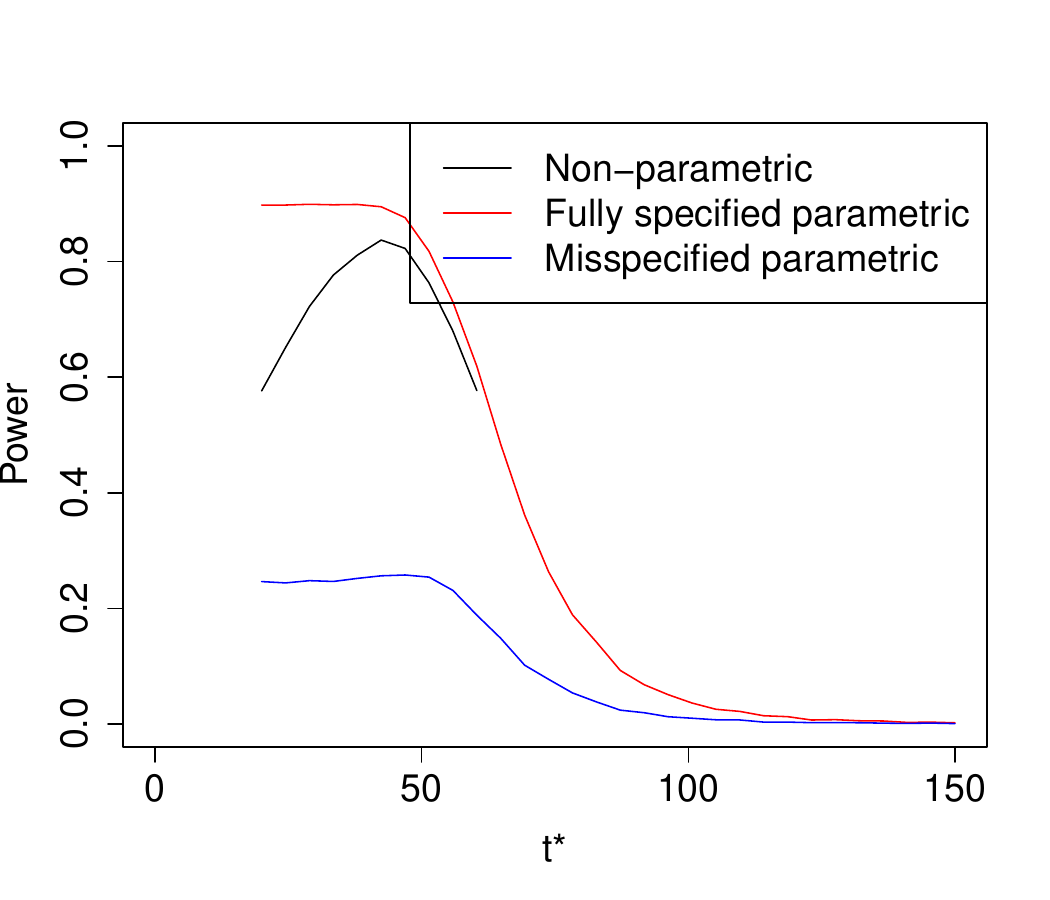}
         \caption{Varying $t^*$ with fixed true knot-point \(t_1=40\) and misspecified knot-point \(\tilde{t}_1=50\).}
            \label{fig:power_crossing_tstar}
     \end{subfigure}
     \hfill
     \begin{subfigure}[t]{0.49\textwidth}
         \centering
         \includegraphics[width=\textwidth]{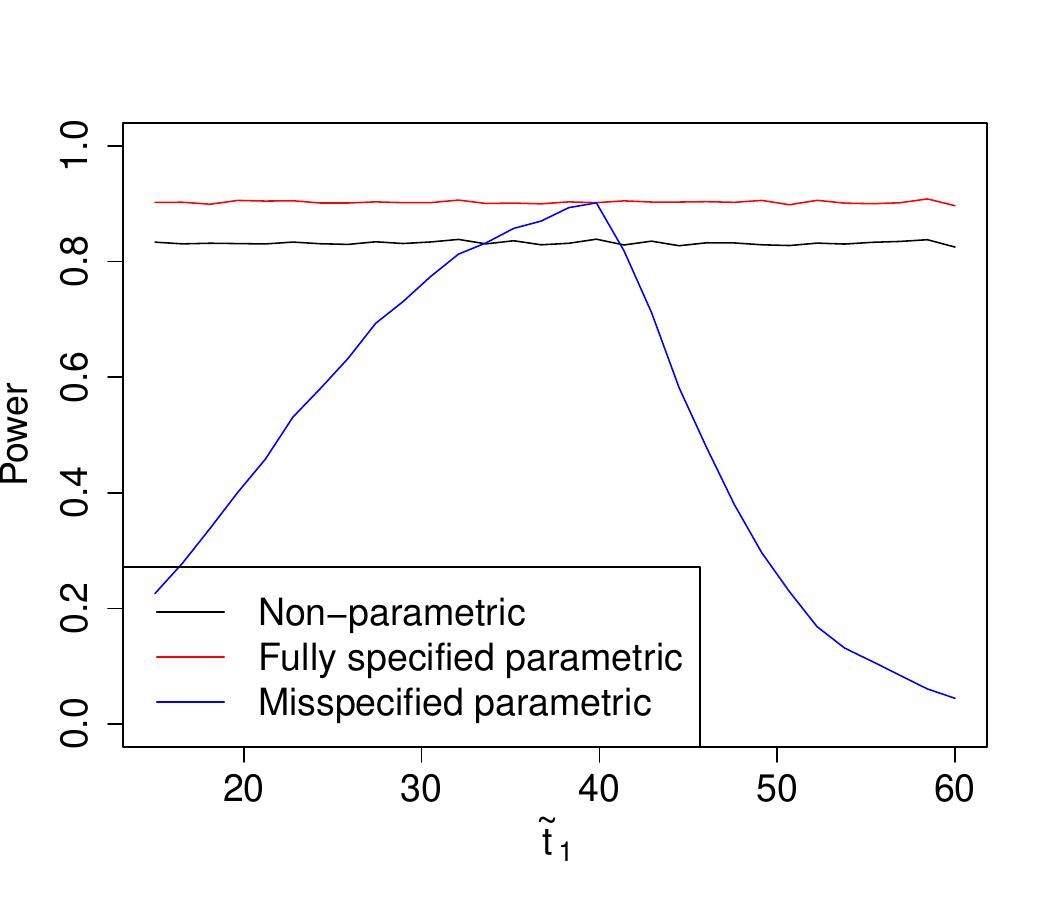}
         \caption{Varying misspecified knot-point \(\tilde{t}_1\) with fixed $t^*=40$ and true knot-point $t_1=40$.}
         \label{fig:power_crossing_t1}
     \end{subfigure}
        \caption{Simulation study results showing power when data is fit to a Cox proportional hazards model with crossing survival curves.}
\end{figure*}

Figure ~\ref{fig:power_crossing_tstar} shows the effect of varying \(t^*\) on power . As expected, for all methods considered, as \(t^*\) increases, power increases up to \(t^*=50\) and decreases beyond \(t^*=50.\) All methods appear highly sensitive to the choice of \(t^*\) with power decreasing rapidly as \(t^*\) increases beyond the time the survival curves cross. The correctly specified fully parametric estimator is most efficient in all cases, reaching power of 90\% compared to maximum power of 80\% for the non-parametric model. Further, the correctly specified fully parametric RMST estimator is identifiable for all \(t^*\) values, unlike the non-parametric estimate which cannot be calculated after the final observed event time in each treatment group. These results compliment the findings of Section~\ref{subsec:measure_comparison}. Figure~\ref{fig:power_crossing_t1} highlights the sensitivity of the parametric RMST estimator to model misspecification. The misspecified parametric RMST estimator increases in efficiency as the model specified knot-point, $\tilde{t}_1$, approaches its true value. An incorrect assumption that the treatment efficacy changes at 20 weeks reduces power to roughly 40\% and an incorrect assumption that efficacy changes at 60 weeks drastically reduces power to almost zero. In fact, it is only more efficient to use a parametric modeling approach over non-parametric if the knot-point is correctly assumed to within 6 weeks of the truth.

In Section \ref{secs5} of the Supplementary material, we discuss the effects of varying \(t^*\) when RMST is used to analyse a trial where the control is initally more effective than the experimental treatment but then becomes less effective after time and the survival curves cross. This is where the curves cross in the opposite direction to what we have discussed here. To address this problem, we simply permute the treatment labels in the simulation. We see that the type I error rates are unaffected but the power is highly susceptible to changes in \(t^*\).

\section{Discussion}\label{sec6}
\subsection{Power and type I error rate for misspecified parametric survival models}
These simulation results demonstrate that the power and type I error rate of misspecified models that utilise the RMST estimator to assess treatment efficacy is dependent on the magnitude of the unknown covariate.  The power of the specified and misspecified estimator coincides as the magnitude of the unknown covariate $\beta_3$ tends to zero. The drop in power appears more significant as $\beta_3$ becomes more negative. The non-parametric estimator follows a similar parabolic relationship with power as the magnitude of $\beta_3$ increases. However, the power of the non-parametric estimator always remains lower than that of the specified estimator. 
In general, whilst the power of the fully specified and misspecified RMST estimator remains stable as the value of $t^*$ varies, the power of the non-parametric RMST estimator is very dependent on $t^*$. The non-parametric estimator only has satisfactory power when $t^*$ is large, towards the end of the trial.

The value of $t^*$ appears to have a generally smaller effect on the type I error rate compared to power, with the non-parametric, misspecified and fully specified estimators all having similar, stable type I error values. It is important to note that these values are all slightly inflated in comparison to the pre-defined error rate due to the generally small sample sizes. 
In comparison, when type I error is evaluated as the unknown covariate varies in magnitude, the misspecified estimator's type I error rate increases significantly. In this case, the type I error rate increases up to 7.3\% as $\beta_3$ becomes more positive and tends towards 2. In contrast, the fully specified and non-parametric estimator's type I error remains unchanged (approximately 3\%) as the size of $\beta_3$ varies.  

\subsection{Power and type I error for crossing curves}
If it is assumed that the underlying data distribution has crossing survival functions, the RMST estimator can be a useful tool for hypothesis testing under suitable choices of parameter values and reasonable prior knowledge of the survival distribution. We see that type I error rates are protected at the 2.5\% level for each method, even when the model is subject to misspecification. This is explained because under \(H_0\), survival curves on different treatments are equal and there is no consequence of modeling these as though they cross over. There is a risk of seeing reduced power if a misspecified parametric RMST estimator is utilised. Misspecification is not an issue for the non-parametric method, however such a method cannot be evaluated for certain values of \(t^*.\) Therefore, the value of \(t^*\) should be clinically meaningful and the timing of the trial analysis should be planned appropriately in accordance with the chosen value of \(t^*.\)

\subsection{Extensions}
Within this paper, simulated survival times and parametric models used to estimate the RMST estimator are Cox-exponential proportional hazard models, with the goal of presenting a best case scenario for power and type I error calculations. Thus, we assume the proportional hazard assumption when assessing efficacy. 
One additional benefit of the non-parametric RMST estimator is that it does not rely on the proportional hazard assumption. Thus, it would be beneficial to consider the impact of interaction terms and crossing survival curves when we assume non-proportional hazards.\cite{ROYST2013} In particular, this work could be expanded to other survival models such as the accelerated failure time model, where the covariate specific hazards associated with survival times are not necessarily constant across time. 

When power is evaluated as $\beta_3$ varied, $t^*$ is chosen at 100 weeks as previous literature has suggested that $t^*$ should be chosen close to the last observed survival time.\cite{ROYST2011} However, research on the RMST estimand is continually evolving.  New literature has presented the Window Mean Survival Time (WMST)\cite{PAUKN2022} which evaluates the mean survival time between an upper and lower time horizon, $\tau_0$ and $\tau_1$ respectively. The WMST can be considered a generalisation of the RMST estimator with $\tau_0=0$ and $\tau_1=t^*$. A further extension to this work could consider power and type I error comparisons for the non-parametric and parametric WMST methods.  

This paper only evaluates the RMST within a fixed sample size trial design. Patients are enrolled to the study during a fixed accrual period and analysis is completed at a fixed time. This work could be extended to evaluate the RMST estimator within an adaptive trial setting, where efficacy is evaluated during interim-analysis time points. This could include the evaluation of power and the type I error rate for the RMST estimator within a group sequential trial design. 

\subsection{Acknowledgements}
We would like to make a special acknowledgement to the MRC Biostatistics Unit's internship programme, which helped make this research possible. DSR received funding from the UK Medical Research Council (MC\_UU\_00002/14). AJB received funding from European Union’s Horizon 2020 research and innovation programme (965397). This research was supported by the NIHR Cambridge Biomedical Research Centre (BRC1215-20014). The views expressed in this publication are those of the authors and not necessarily those of the NHS, the National Institute for Health Research or the Department of Health and Social Care (DHCS). For the purpose of open access, the author has applied a Creative Commons Attribution (CC BY) licence to any Author Accepted Manuscript version arising.

\subsection{Data availability statement}
The code used to simulate results presented in this paper is publicly available in the following GitHub repository: \url{https://github.com/alemily100/rmst-simulations}

\subsection{Orcid}
Emily Alger: 0000-0002-5378-7439 \\
David S.\ Robertson: 0000-0001-6207-0416 \\
Abigail J.\ Burdon: 0000-0002-0883-4160

\printbibheading[title={References},heading=subbibnumbered]
\printbibliography[heading=none]


\clearpage

\section{Supplementary material} \label{sec7}
\subsection{Deriving the fully specified RMST estimator}\label{secs1}
Within this case study, the covariates $X_\text{Treatment}$, $X_\text{Inherit}$ and $X_\text{sex}$ are generated using Bernoulli(0.5) random variables. In the following section we use the indicator function  to show the presence of each covariate. For example, $\mathbbm{1}_{X_\text{Treatment}} \in \{0,1\}$ and is equal to 1 if a patient is on the treatment arm and 0 if the patient is on the control arm. The interaction between `treatment' and `inherit' is defined as $X_\text{Treatment:Inherit} = \min(X_\text{Treatment}, X_\text{Inherit})$. Table 1 in the main manuscript presents the associated values of each indicator function for all covariates. 
\subsubsection{Treatment arm}
The instances, and probability of occurrence, for each possible combination of covariates for patients in  the treatment arm is presented in Table \ref{tbl:occurences}.

Given $X_\text{Treatment}=1$,
\begin{table}[hbt!]
\centering
\begin{tabular}{c|c|c|c}
 $\mathbbm{1}_{X_\text{Inherit}}$ & $\mathbbm{1}_{X_\text{Treatment:Inherit}}$ & $\mathbbm{1}_{X_\text{Sex}}$ & \textbf{Probability of occurrence} \\ \hline
 1 & 1 & 1 & 0.25 \\
 1 & 1 & 0 & 0.25 \\
 0 & 0 & 1 & 0.25 \\
 0 & 0 & 0 & 0.25
\end{tabular}
\caption{All combinations of covariate indicators for patients in the treatment arm, reported alongside probability of occurrence.}
\label{tbl:occurences}
\end{table}

Hence, the associated corresponding survival function $S_1(t)$ for patients in the treatment arm with vector of fitted parameter coefficients $\boldsymbol{\hat{\theta}}$ and fitted base hazard $\hat{\lambda}_1$ is,
\begin{multline*}
    \hat{S}_1(t) = 0.25 \exp(-\hat{\lambda}_1 t \exp(\hat{\theta}_1+\hat{\theta}_2+\hat{\theta}_{12}+\hat{\theta}_3)) + 0.25 \exp(-\hat{\lambda}_1 t \exp(\hat{\theta}_1+\hat{\theta}_2+\hat{\theta}_{12})) + \\ + 0.25 \exp(-\hat{\lambda}_1 t \exp(\hat{\theta}_1+\hat{\theta}_3)) + 0.25 \exp(-\hat{\lambda}_1 t \exp(\hat{\theta}_1)).
\end{multline*}

\clearpage

\subsubsection{Control arm}

The instances, and probability of occurrence, for each possible combination of covariates for patients in the control arm is presented in Table \ref{tbl:occurences_control}.

Given $X_\text{Treatment}=0$,
\begin{table}[htb!]
\centering
\begin{tabular}{c|c|c|c}
 $\mathbbm{1}_{X_\text{Inherit}}$ & $\mathbbm{1}_{X_\text{Treatment:Inherit}}$ & $\mathbbm{1}_{X_\text{Sex}}$ & \textbf{Probability of occurrence} \\ \hline
 1 & 0 & 1 & 0.25 \\
 1 & 0 & 0 & 0.25 \\
 0 & 0 & 1 & 0.25 \\
 0 & 0 & 0 & 0.25
\end{tabular}
\caption{All combinations of covariate indicators for patients in the control arm, reported alongside probability of occurrence.}
\label{tbl:occurences_control}
\end{table}

Similarly, the associated survival function under the control arm is, 
\begin{multline*}
    \hat{S}_0(t)= 0.25 \exp (-\hat{\lambda}_1 t \exp(\hat{\theta}_2 + \hat{\theta}_3)) + 0.25 \exp(-\hat{\lambda}_1 t \exp(\hat{\theta}_2)) + 0.25 \exp(-\hat{\lambda}_1 t \exp(\hat{\theta}_3)) + 0.25 \exp(-\hat{\lambda}_1 t).
\end{multline*}

  In this instance, 

  \begin{align*}
      \hat{\Delta}_\text{full}(t^*|\hat{\boldsymbol{\theta}}, \hat{\lambda}_1) &= \hat{\mu}_\text{full}(t^*|\hat{\boldsymbol{\theta}},\hat{\lambda}_1,X_\text{Treatment}=1)-\hat{\mu}_\text{full}(t^*|\hat{\boldsymbol{\theta}},\hat{\lambda}_1,X_\text{Treatment}=0), \\
      &=  \int_0^{t^*} \hat{S}_1(t) \;dt  - \int_0^{t^*} \hat{S}_0(t) \;dt.
  \end{align*}

  A similar approach is used to estimate the mis-specified parametric RMST. 

\newpage

\FloatBarrier
\subsection{Checking normality of Z-statistics}\label{secs2}
\begin{figure*}[!htb]
     \centering
     \begin{subfigure}[t]{0.49\textwidth}
         \centering
         \includegraphics[width=\textwidth, height=7.5cm]{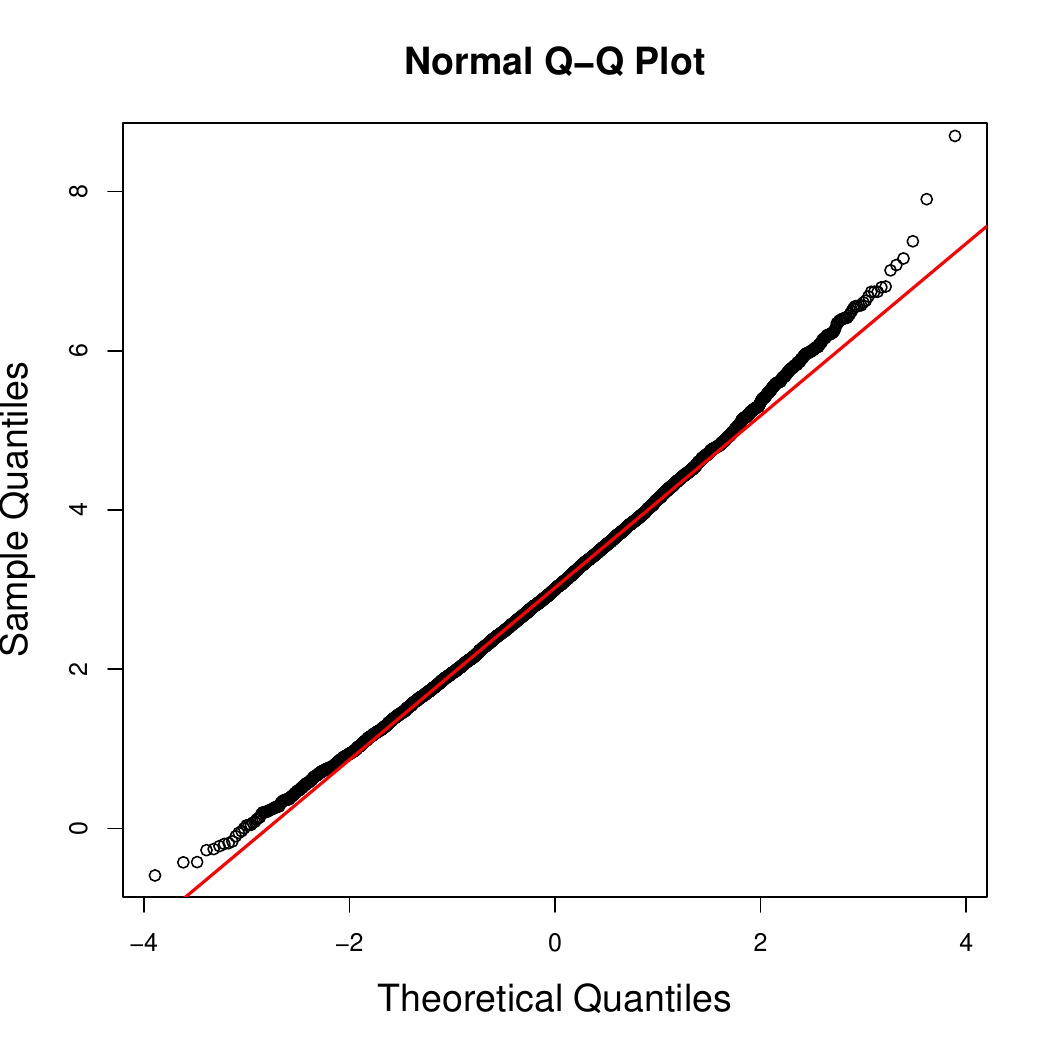}
         \caption{Q-Q plot of Z-statistics.}
     \end{subfigure}
     \hfill
     \begin{subfigure}[t]{0.49\textwidth}
         \centering
         \includegraphics[width=\textwidth, height=7.5cm]{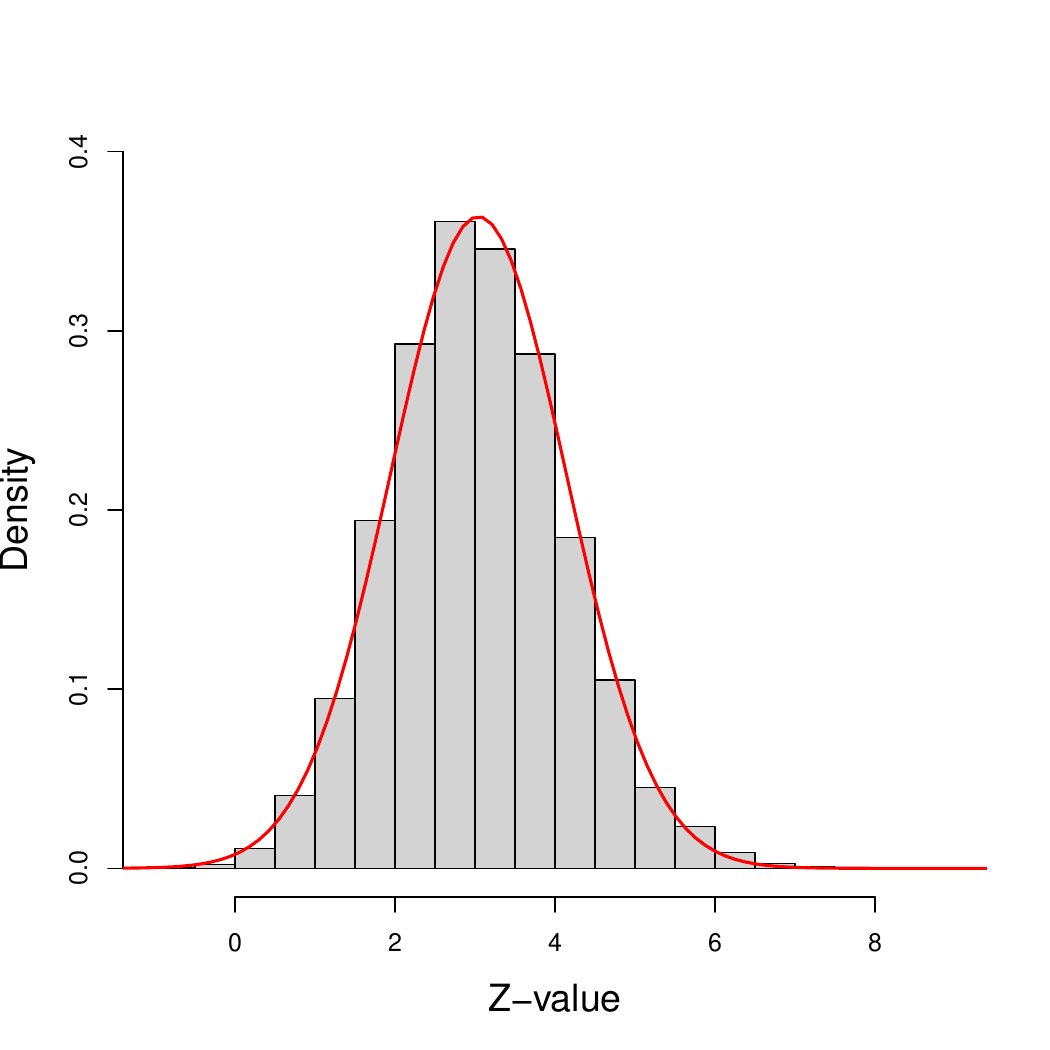}
         \caption{Histogram of Z-statistics with a normal distribution overlayed using sample mean and sample standard deviation.}
     \end{subfigure}
        \caption{Graphs produced to assess normality for $N=10^4$ Z-statistics using the non-parametric RMST estimator.}
\end{figure*}

\begin{figure*}[!htb]
     \centering
     \begin{subfigure}[t]{0.49\textwidth}
         \centering
         \includegraphics[width=\textwidth, height=7.5cm]{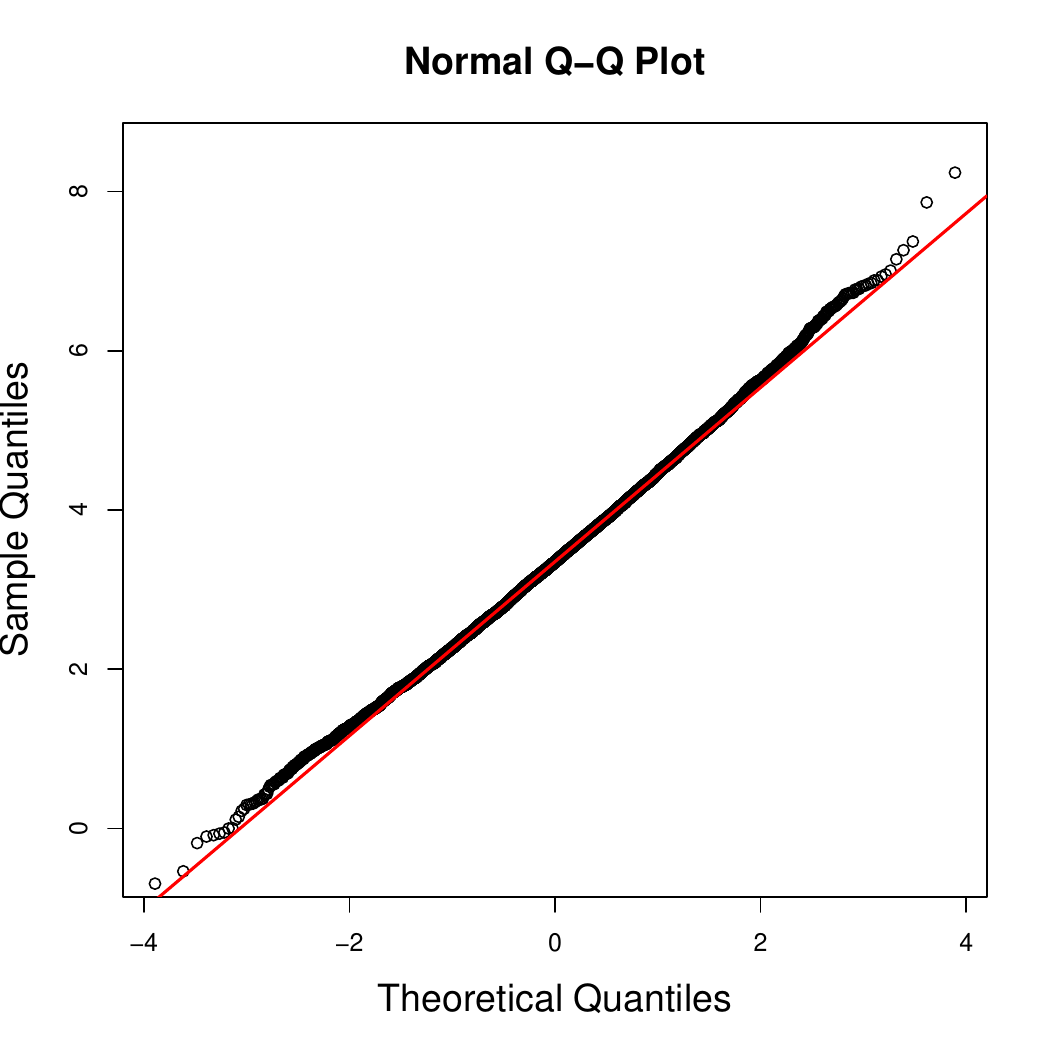}
         \caption{Q-Q plot of Z-statistics.}
     \end{subfigure}
     \hfill
     \begin{subfigure}[t]{0.49\textwidth}
         \centering
         \includegraphics[width=\textwidth, height=7.5cm]{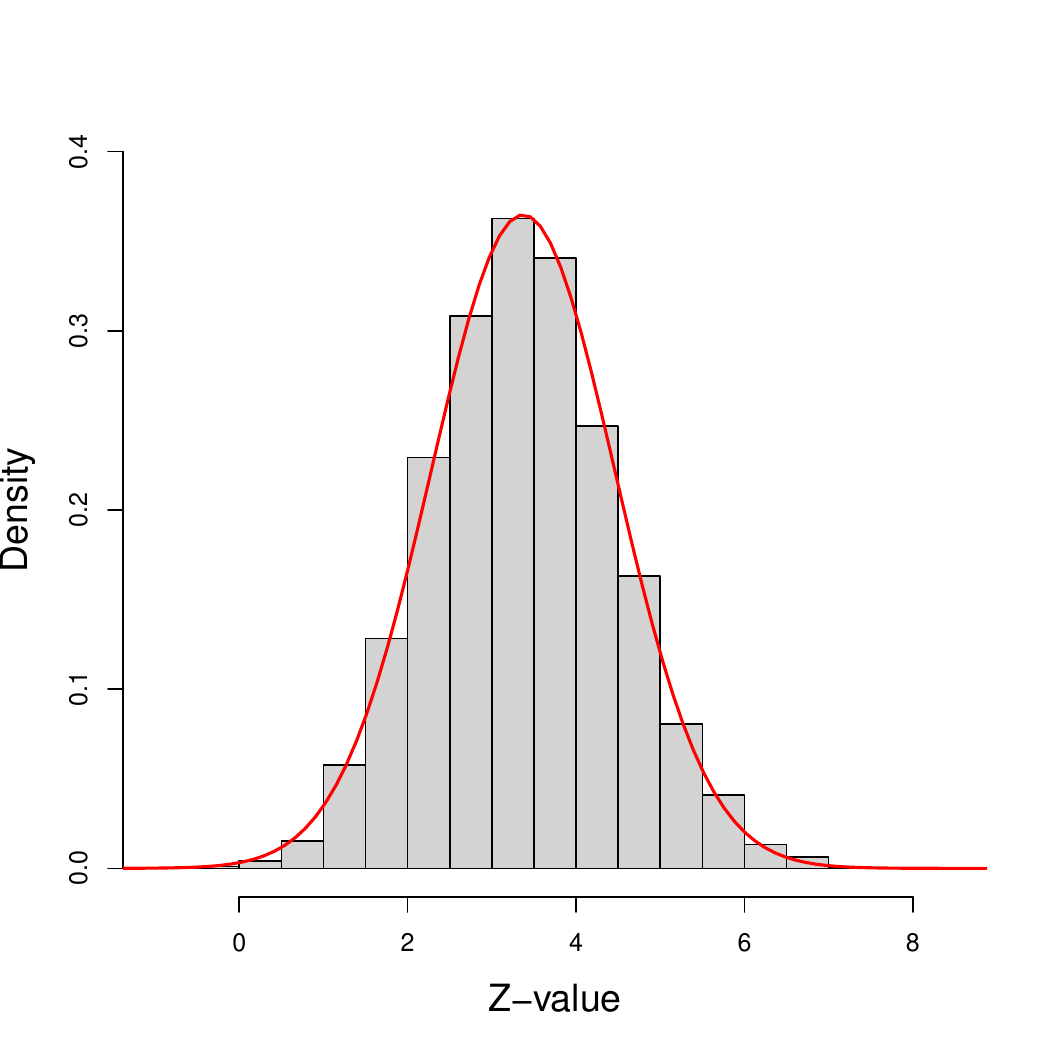}
         \caption{Histogram of Z-statistics with a normal distribution overlayed using sample mean and sample standard deviation.}
     \end{subfigure}
        \caption{Graphs produced to assess normality for $N=10^4$ Z-statistics using the fully specified parametric RMST estimator.}
\end{figure*}

\begin{figure*}[!htb]
     \centering
     \begin{subfigure}[t]{0.49\textwidth}
         \centering
         \includegraphics[width=\textwidth, height=7.5cm]{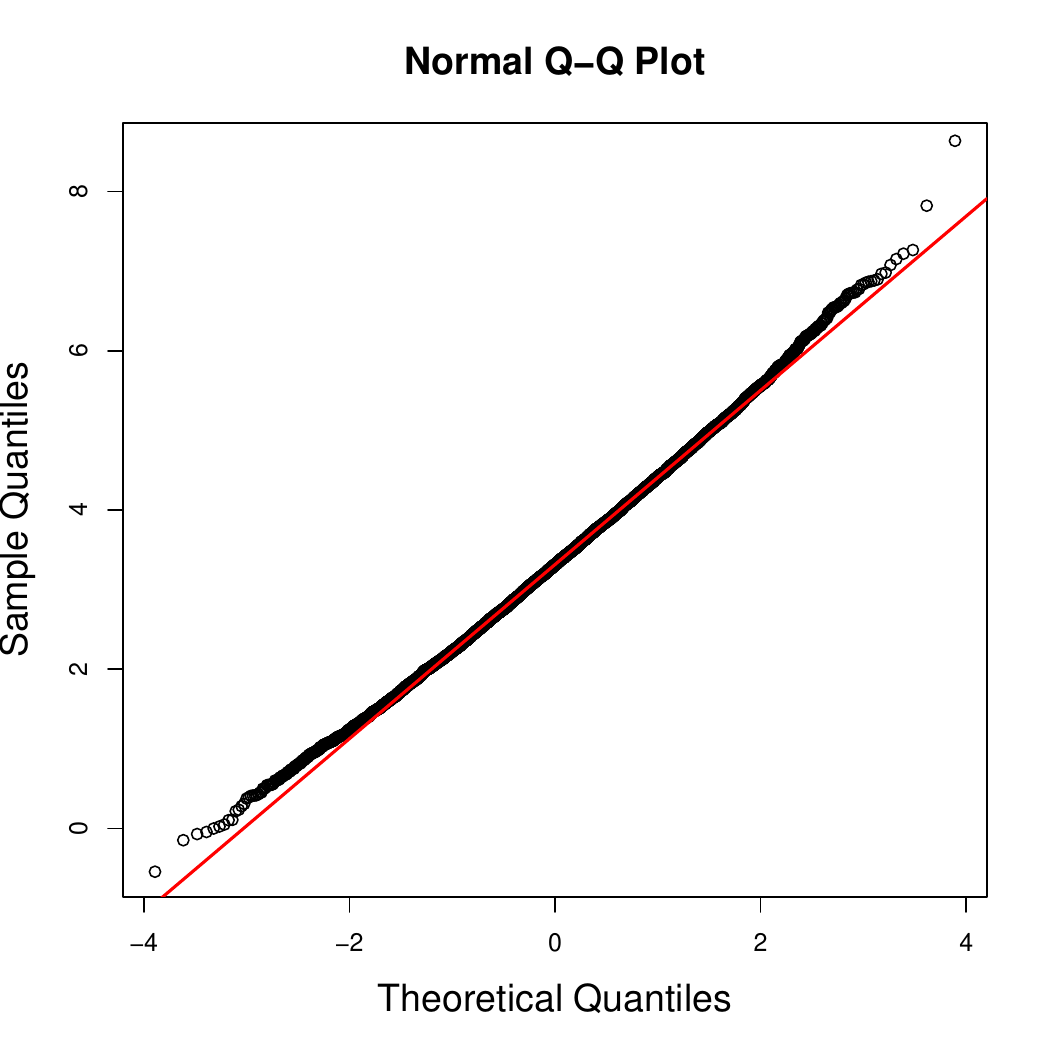}
         \caption{Q-Q plot of Z-statistics.}
     \end{subfigure}
     \hfill
     \begin{subfigure}[t]{0.49\textwidth}
         \centering
         \includegraphics[width=\textwidth, height=7.5cm]{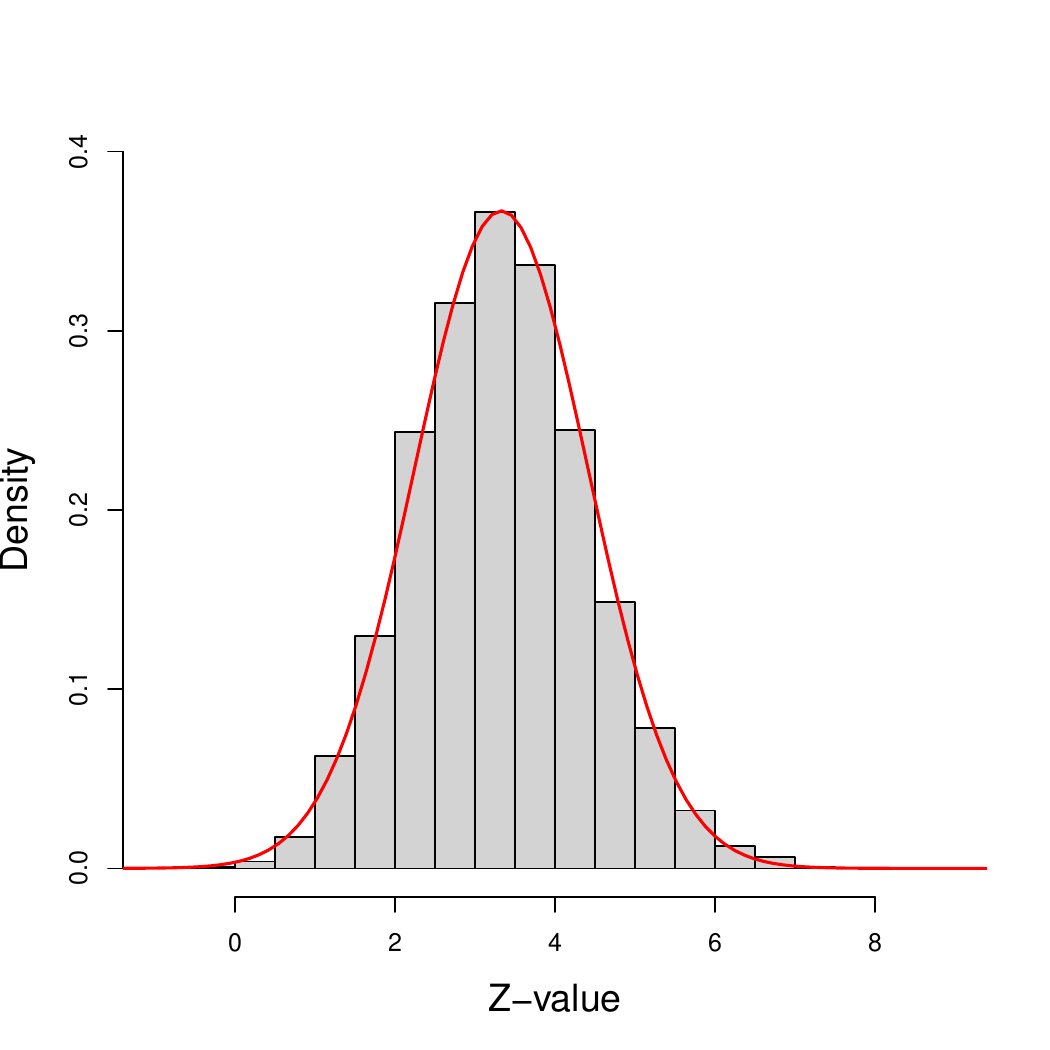}
         \caption{Histogram of Z-statistics with a normal distribution overlayed using sample mean and sample standard deviation.}
     \end{subfigure}
        \caption{Graphs produced to assess normality for $N=10^4$ Z-statistics using the misspecified parametric RMST estimator.}
\end{figure*}

\clearpage

\FloatBarrier
\subsection{Tabulated Z-value breakdown}\label{secs3}
Relevant to Section 4 of the main manuscript, below we present a breakdown of mean standard error, treatment effect, and Z-statistics used to assess power and type I error as $t^*$ and $\beta_3$ varied. 
\begin{table}[!htb]
\begin{adjustwidth}{-.5in}{-.5in}
\label{tbl:nonparam_power}
\begin{center}
\begin{tabular}{ccccc|cccc|cccc}
 &
  \multicolumn{4}{c|}{\bf{Non-parametric estimator}} &
  \multicolumn{4}{c|}{\bf{Fully specified parametric estimator}} &
  \multicolumn{4}{c}{\bf{Mis-specified parametric estimator}} \\ \cline{2-13} 
$\boldsymbol{\beta_3}$ &
  $Z$ &
  $\text{s.e}(\hat{\Delta}(\beta_3))$ &
  $\hat{\Delta}(\beta_3)$ &
  Power &
  $Z$ &
  $\text{s.e}(\hat{\Delta}(\beta_3))$ &
  $\hat{\Delta}(\beta_3)$ &
  Power &
  $Z$ &
  $\text{s.e}(\hat{\Delta}(\beta_3))$ &
  $\hat{\Delta}(\beta_3)$ &
  Power \\$\mathbf{-2.00}$&2.16&6.67&14.31&0.57&2.78&5.24&14.40&0.77&2.26&5.77&12.97&0.60\\$\mathbf{-1.56}$&2.35&6.75&15.71&0.64&2.89&5.51&15.79&0.80&2.50&5.92&14.72&0.69\\$\mathbf{-1.11}$&2.58&6.85&17.53&0.72&3.05&5.83&17.64&0.84&2.80&6.09&16.95&0.78\\$\mathbf{-0.67}$&2.87&6.93&19.69&0.80&3.24&6.15&19.80&0.88&3.14&6.26&19.50&0.86\\$\mathbf{-0.22}$&3.18&6.98&21.98&0.87&3.48&6.38&22.07&0.92&3.48&6.38&22.01&0.92\\$\mathbf{0.22}$&3.43&6.97&23.66&0.91&3.72&6.42&23.72&0.95&3.71&6.42&23.67&0.95\\$\mathbf{0.67}$&3.50&6.93&23.97&0.92&3.90&6.20&23.98&0.96&3.77&6.39&23.88&0.95\\$\mathbf{1.11}$&3.29&6.93&22.58&0.89&3.94&5.74&22.54&0.97&3.63&6.32&22.70&0.92\\$\mathbf{1.56}$&2.89&7.01&20.06&0.81&3.90&5.14&20.02&0.97&3.39&6.23&20.91&0.87\\$\mathbf{2.00}$&2.45&7.14&17.39&0.68&3.85&4.51&17.38&0.97&3.18&6.14&19.30&0.82\\
\end{tabular}
\caption{Mean simulated Z-statistic, treatment effect, and associated standard error to assess power for each RMST estimator as $\beta_3$ varied for $N=10^5$ simulations.}
\end{center}
\end{adjustwidth}
\end{table}
\begin{table}[!htb]
\begin{adjustwidth}{-.5in}{-.5in}
\label{tbl:nonparam_power_tstar}
\begin{center}
\begin{tabular}{ccccc|cccc|cccc}
 &
  \multicolumn{4}{c|}{\bf{Non-parametric estimator}} &
  \multicolumn{4}{c|}{\bf{Fully specified parametric estimator}} &
  \multicolumn{4}{c}{\bf{Mis-specified parametric estimator}} \\ \cline{2-13} 
$\boldsymbol{t^*}$ &
  $Z$ &
  $\hat{\Delta}(t^*)$ &
  $\text{s.e}(\hat{\Delta}(t^*))$ &
  Power &
  $Z$ &
  $\hat{\Delta}(t^*)$ &
  $\text{s.e}(\hat{\Delta}(t^*))$ &
  Power &
  $Z$ &
  $\hat{\Delta}(t^*)$ &
  $\text{s.e}(\hat{\Delta}(t^*))$ &
  Power\\$\mathbf{10.71}$&1.12&0.37&0.42&0.21&2.99&0.15&0.44&0.88&2.98&0.14&0.42&0.88\\$\mathbf{21.43}$&1.57&1.00&1.56&0.35&3.06&0.53&1.62&0.88&3.04&0.51&1.57&0.88\\$\mathbf{32.14}$&1.90&1.74&3.28&0.47&3.12&1.08&3.38&0.89&3.09&1.06&3.29&0.88\\$\mathbf{42.86}$&2.16&2.54&5.44&0.57&3.18&1.75&5.58&0.89&3.14&1.73&5.46&0.89\\$\mathbf{53.57}$&2.39&3.37&7.97&0.65&3.24&2.52&8.14&0.90&3.19&2.50&7.98&0.89\\$\mathbf{64.29}$&2.57&4.21&10.74&0.71&3.28&3.34&10.92&0.90&3.23&3.33&10.75&0.89\\$\mathbf{75.00}$&2.74&5.05&13.73&0.76&3.32&4.20&13.9&0.90&3.28&4.20&13.72&0.90\\$\mathbf{85.71}$&2.89&5.88&16.83&0.80&3.35&5.09&16.98&0.90&3.31&5.11&16.81&0.90\\$\mathbf{96.43}$&3.02&6.70&20.00&0.83&3.38&6.00&20.12&0.91&3.34&6.03&19.98&0.90\\$\mathbf{107.14}$&3.13&7.50&23.24&0.86&3.40&6.91&23.32&0.91&3.36&6.96&23.21&0.90\\$\mathbf{117.86}$&3.14&8.31&25.84&0.86&3.42&7.83&26.52&0.91&3.38&7.89&26.44&0.90\\$\mathbf{128.57}$&NA&NA&NA&NA&3.43&8.74&29.69&0.91&3.39&8.82&29.65&0.90\\$\mathbf{139.29}$&NA&NA&NA&NA&3.44&9.65&32.87&0.91&3.41&9.74&32.86&0.91\\$\mathbf{150.00}$&NA&NA&NA&NA&3.44&10.54&35.96&0.91&3.41&10.65&35.99&0.91\\
\end{tabular}
\caption{Mean simulated Z-statistic, treatment effect, and associated standard error to assess power for each RMST estimator as $t^* $ varied for 100,000 simulations. 
NA values shown when non-parametric RMST could not be calculated as $t^*$ was beyond the end of the trial (120 weeks).}
\end{center}
\end{adjustwidth}
\end{table}
\begin{table}
\begin{adjustwidth}{-.5in}{-.5in}
\label{tbl:nonparam_power_type1}
\begin{center}
\begin{tabular}{ccccc|cccc|cccc}
 &
  \multicolumn{4}{c|}{\bf{Non-parametric estimator}} &
  \multicolumn{4}{c|}{\bf{Fully specified parametric estimator}} &
  \multicolumn{4}{c}{\bf{Mis-specified parametric estimator}} \\ \cline{2-13} 
$\boldsymbol{\beta_3}$ &
  $Z$ &
  $\text{s.e}(\hat{\Delta}(\beta_3))$ &
  $\hat{\Delta}(\beta_3)$ &
  TIE &
  $Z$ &
  $\text{s.e}(\hat{\Delta}(\beta_3))$ &
  $\hat{\Delta}(\beta_3)$ &
  TIE &
  $Z$ &
  $\text{s.e}(\hat{\Delta}(\beta_3))$ &
  $\hat{\Delta}(\beta_3)$ &
  TIE \\$\mathbf{-2.00}$&0.00&7.30&0.02&0.029&0.00&5.57&0.01&0.030&0.00&6.36&0.020&0.034\\$\mathbf{-1.56}$&0.00&7.34&0.00&0.028&0.00&5.89&0.00&0.030&0.00&6.51&0.00&0.033\\$\mathbf{-1.11}$&0.00&7.36&0.01&0.028&0.00&6.25&0.02&0.030&0.00&6.65&0.03&0.032\\$\mathbf{-0.67}$&0.00&7.31&0.02&0.028&0.00&6.57&0.00&0.029&0.00&6.74&0.01&0.030\\$\mathbf{-0.22}$&0.00&7.13&0.00&0.028&0.00&6.70&-0.02&0.029&0.00&6.70&-0.01&0.029\\$\mathbf{0.22}$&0.00&6.84&-0.01&0.027&0.00&6.51&0.00&0.028&0.00&6.52&0.00&0.028\\$\mathbf{0.67}$&0.00&6.53&0.01&0.028&0.00&5.97&0.00&0.028&0.00&6.23&0.01&0.033\\$\mathbf{1.11}$&0.00&6.37&0.02&0.026&0.00&5.23&0.01&0.026&0.00&5.93&0.030&0.043\\$\mathbf{1.56}$&0.00&6.39&0.00&0.026&0.01&4.55&0.02&0.027&0.00&5.67&0.01&0.059\\$\mathbf{2.00}$&0.00&6.51&-0.01&0.025&0.00&4.05&-0.02&0.027&0.00&5.46&-0.01&0.073\\
\end{tabular}
\caption{Mean simulated Z-statistic, treatment effect, and associated standard error to assess Type I error (TIE) rate for each RMST estimator as $\beta_3 $ varied for $N=10^5$ simulations.}
\end{center}
\end{adjustwidth}
\end{table}
\begin{table}
\begin{adjustwidth}{-.5in}{-.5in}
\label{tbl:nonparam_power_tstar}
\begin{center}
\begin{tabular}{ccccc|cccc|cccc}
 &
  \multicolumn{4}{c|}{\bf{Non-parametric estimator}} &
  \multicolumn{4}{c|}{\bf{Fully specified parametric estimator}} &
  \multicolumn{4}{c}{\bf{Mis-specified parametric estimator}} \\ \cline{2-13} 
$\boldsymbol{t^*}$ &
  $Z$ &
  $\text{s.e}(\hat{\Delta}(t^*))$ &
  $\hat{\Delta}(t^*)$ &
  TIE &
  $Z$ &
  $\text{s.e}(\hat{\Delta}(t^*))$ &
  $\hat{\Delta}(t^*)$ &
  TIE &
  $Z$ &
  $\text{s.e}(\hat{\Delta}(t^*))$ &
  $\hat{\Delta}(t^*)$ &
  TIE \\$\mathbf{10.71}$&0.001&0.434&0.001&0.028&0.001&0.177&0.000&0.025&0.001&0.171&0.000&0.025\\$\mathbf{21.43}$&0.003&1.153&0.003&0.028&0.004&0.632&0.002&0.026&0.004&0.617&0.002&0.026\\$\mathbf{32.14}$&0.000&1.980&0.001&0.028&0.002&1.276&0.002&0.028&0.002&1.255&0.002&0.028\\$\mathbf{42.86}$&0.001&2.851&0.002&0.027&0.000&2.043&0.001&0.028&0.001&2.021&0.003&0.027\\$\mathbf{53.57}$&0.003&3.727&0.011&0.028&0.000&2.880&-0.001&0.029&-0.001&2.864&-0.004&0.029\\$\mathbf{64.29}$&-0.002&4.590&-0.009&0.027&0.001&3.757&0.003&0.028&0.001&3.753&0.003&0.028\\$\mathbf{75.00}$&0.002&5.423&0.013&0.027&0.001&4.646&0.006&0.029&0.000&4.659&0.001&0.029\\$\mathbf{85.71}$&0.004&6.220&0.025&0.027&0.003&5.529&0.015&0.029&0.003&5.561&0.015&0.029\\$\mathbf{96.43}$&0.001&6.976&0.004&0.028&0.002&6.393&0.016&0.029&0.002&6.446&0.015&0.030\\$\mathbf{107.14}$&0.004&7.689&0.028&0.027&0.003&7.228&0.020&0.028&0.002&7.303&0.016&0.028\\$\mathbf{117.86}$&0.000&8.420&-0.003&0.025&0.002&8.030&0.017&0.029&0.002&8.128&0.016&0.029\\$\mathbf{120.00}$&NA&NA&NA&NA&0.007&8.188&0.061&0.029&0.008&8.288&0.065&0.030\\$\mathbf{124.00}$&NA&NA&NA&NA&0.001&8.476&0.008&0.029&0.001&8.583&0.004&0.029\\$\mathbf{128.00}$&NA&NA&NA&NA&-0.004&8.757&-0.033&0.029&-0.003&8.872&-0.030&0.029\\$\mathbf{132.00}$&NA&NA&NA&NA&0.004&9.033&0.037&0.028&0.003&9.154&0.026&0.029\\$\mathbf{136.00}$&NA&NA&NA&NA&-0.003&9.304&-0.028&0.028&-0.004&9.432&-0.036&0.028\\$\mathbf{140.00}$&NA&NA&NA&NA&0.000&9.567&0.005&0.029&-0.001&9.703&-0.008&0.029\\$\mathbf{144.00}$&NA&NA&NA&NA&0.005&9.826&0.048&0.029&0.005&9.966&0.051&0.029\\$\mathbf{148.00}$&NA&NA&NA&NA&0.000&10.082&-0.001&0.028&-0.001&10.229&-0.012&0.028\\
\end{tabular}
\end{center}
\caption{Mean simulated Z-statistic, treatment effect, and associated standard error to assess Type I error (TIE) rate for each RMST estimator as $t^* $ varied for $N=10^5$ simulations. 
NA values shown when non-parametric RMST could not be calculated as $t^*$ was beyond the end of the trial (120 weeks).}
\end{adjustwidth}
\end{table}

\clearpage

\FloatBarrier
\subsection{Graphical Z-value breakdown}\label{secs4}
Relevant to Section 4 of the main manuscript, below we present figures of the mean standard error, treatment effect, and Z-statistics used to assess power and type I error as $t^*$ and $\beta_3$ varied. 

\begin{figure}[ht!]
\centering
\begin{subfigure}{0.45\textwidth}
    \includegraphics[width=\textwidth]{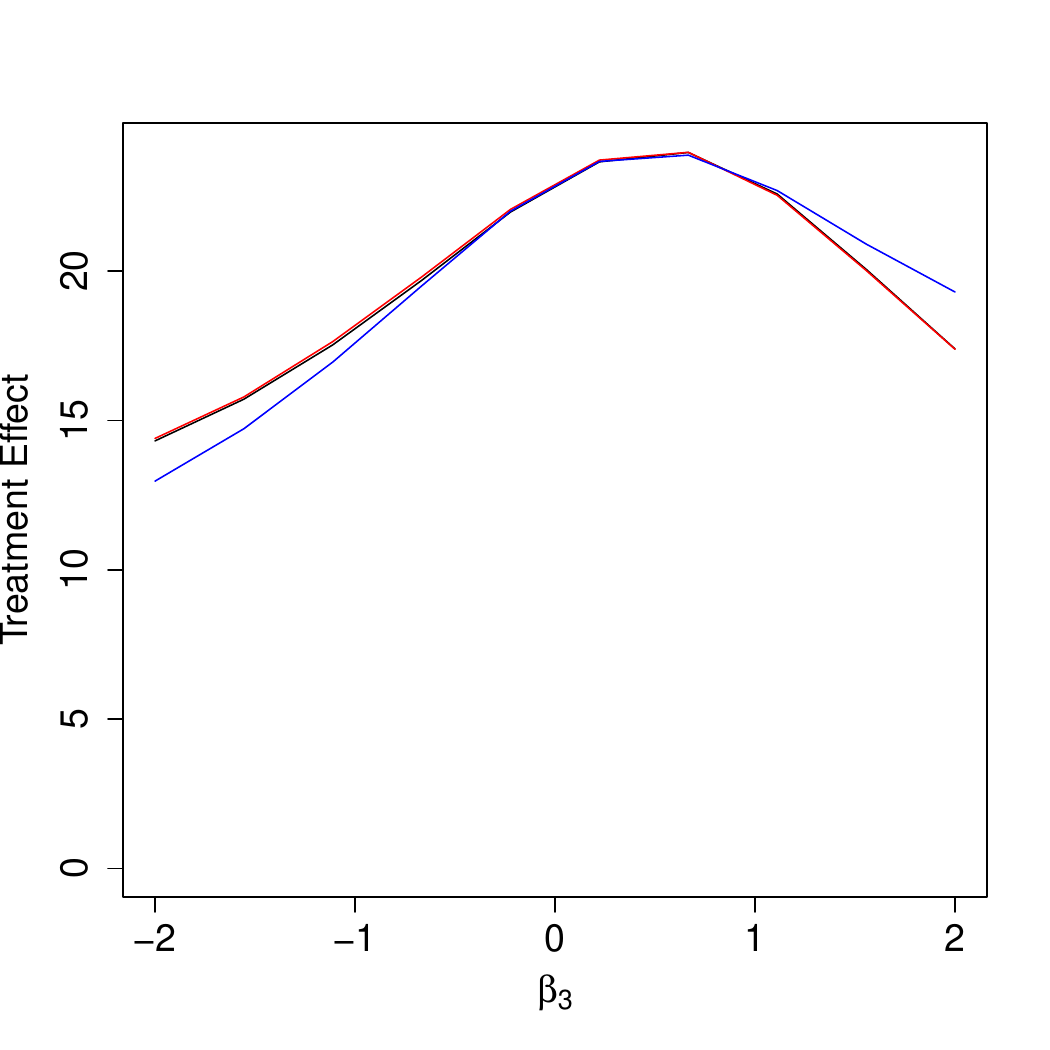}
    \caption{Mean treatment effect for varying $\beta_3$.}
    \label{fig:first}
\end{subfigure}
\hfill
\begin{subfigure}{0.45\textwidth}
    \includegraphics[width=\textwidth]{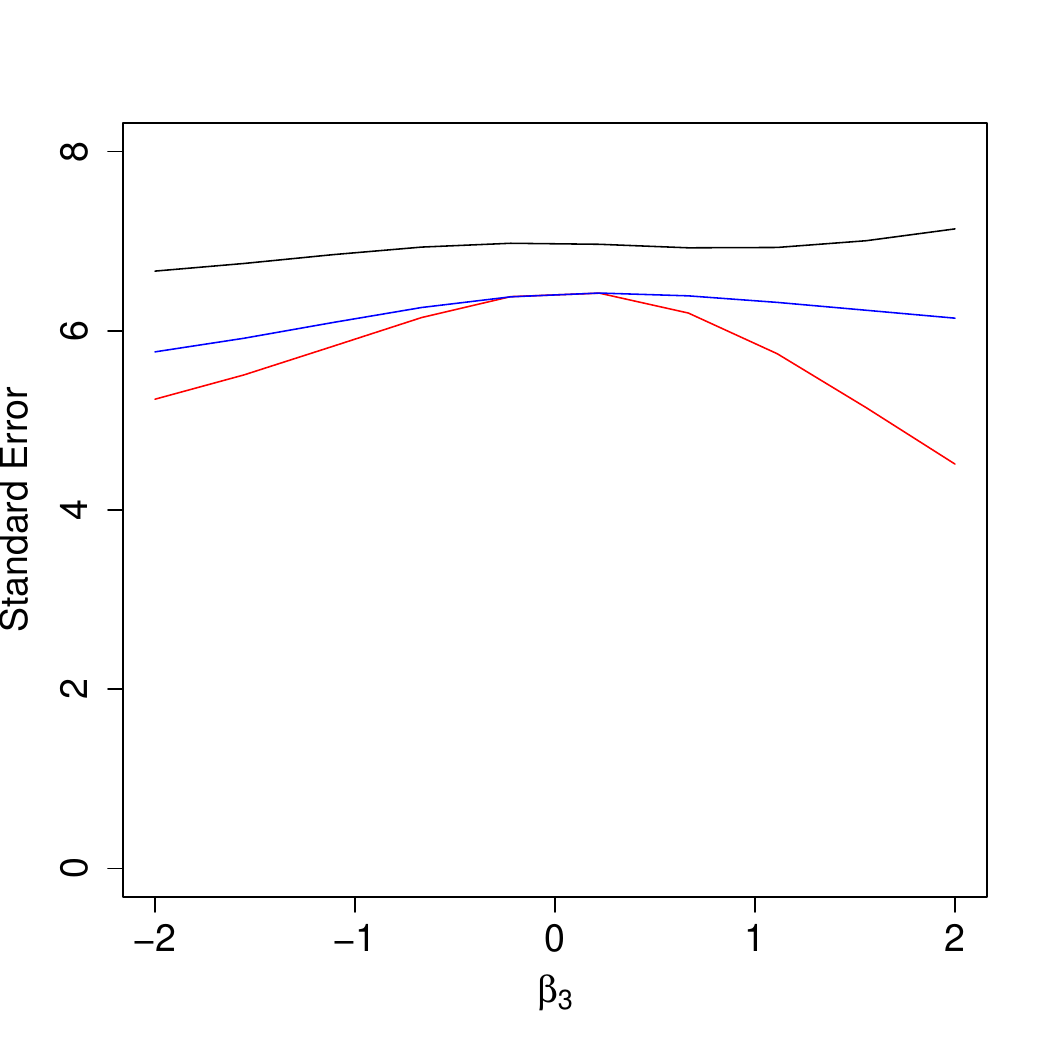}
    \caption{Mean standard error for varying $\beta_3$.}
    \label{fig:second}
\end{subfigure}
\hfill
\begin{subfigure}{\textwidth}
    \includegraphics[width=\textwidth]{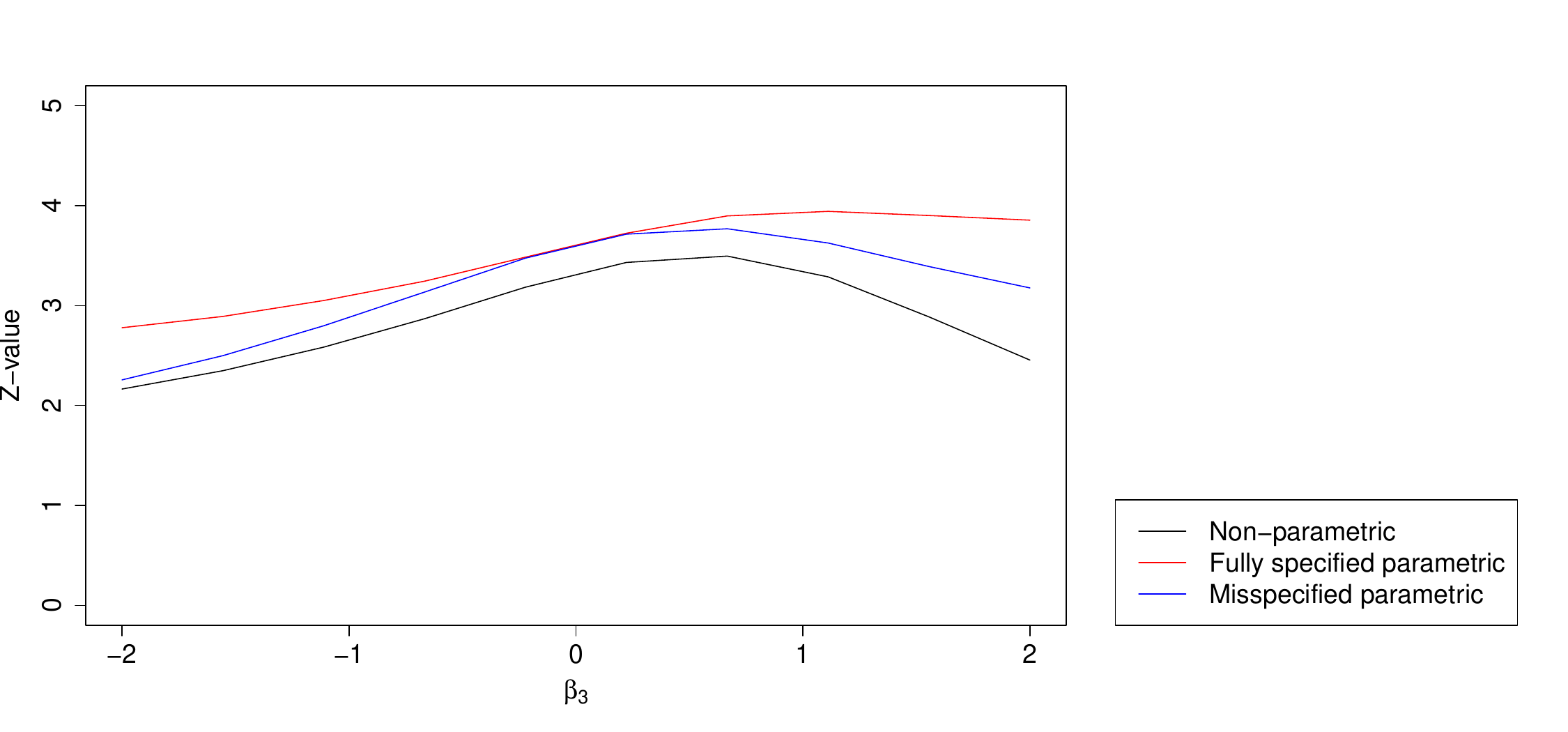}
    \caption{Mean Z-statistic for varying $\beta_3$.}
    \label{fig:third}
\end{subfigure}
\caption{Power breakdown for varying $\beta_3$ for each RMST estimator for $N=10^5$ simulations.}
\end{figure}

\begin{figure}[ht!]
\centering
\begin{subfigure}{0.45\textwidth}
    \includegraphics[width=\textwidth]{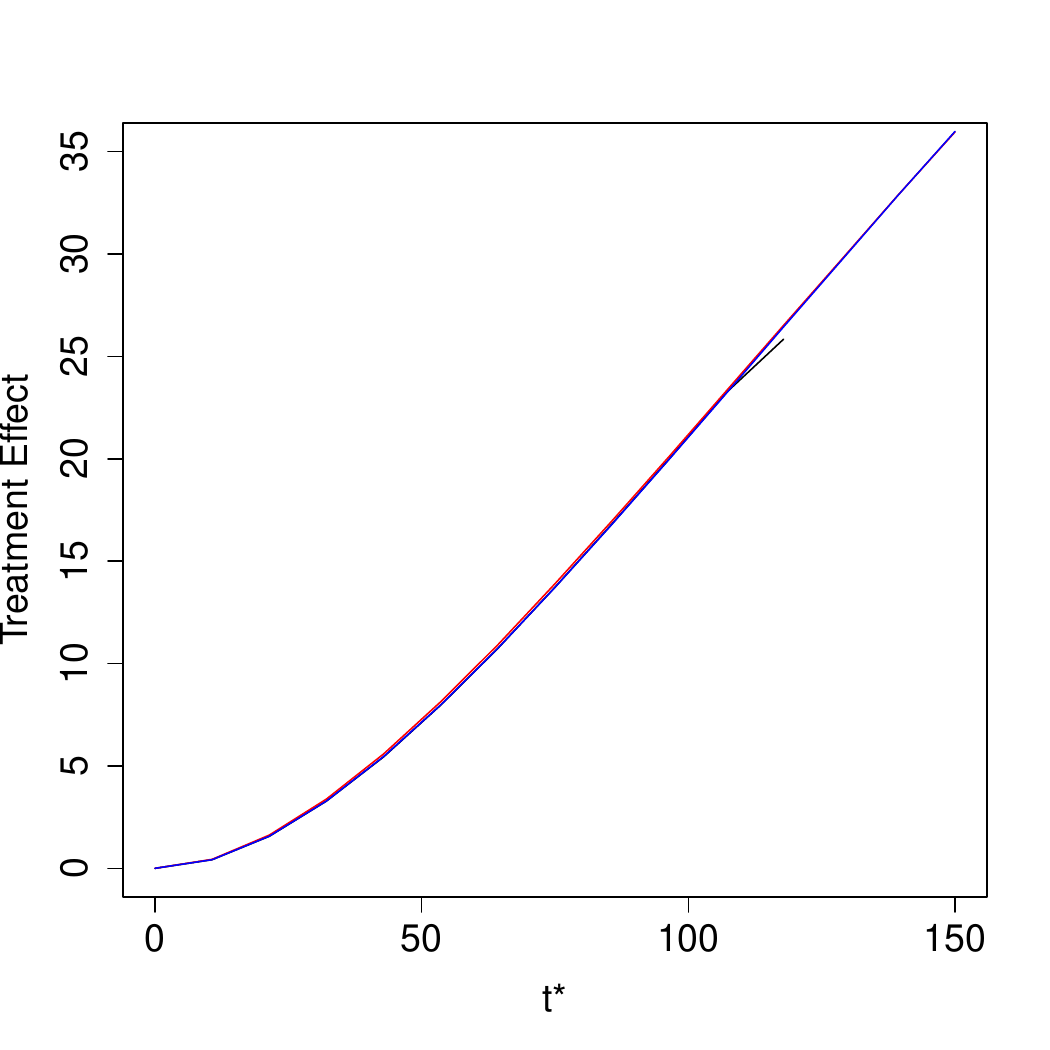}
    \caption{Mean treatment effect for varying $t^*$.}
    \label{fig:first}
\end{subfigure}
\hfill
\begin{subfigure}{0.45\textwidth}
    \includegraphics[width=\textwidth]{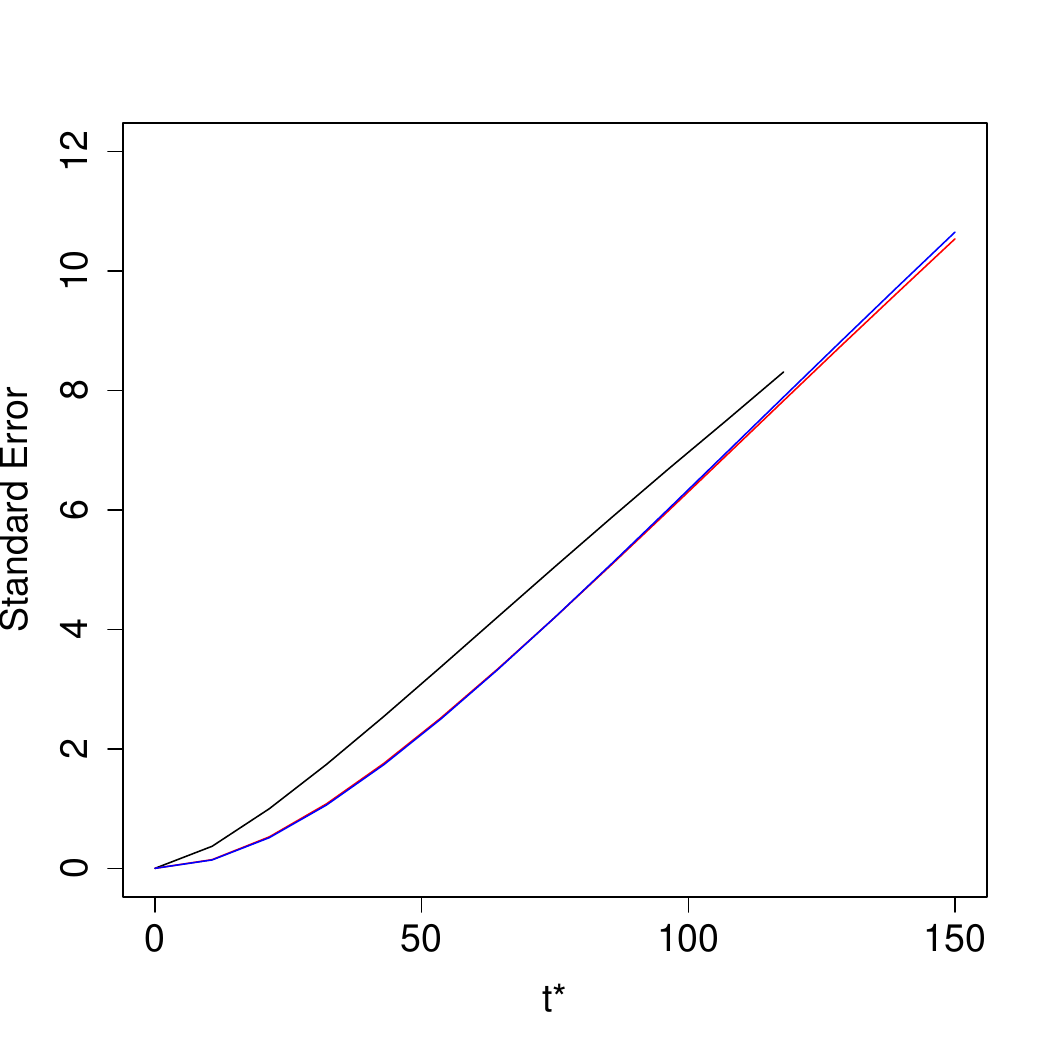}
    \caption{Mean standard error for varying $\beta_3$.}
    \label{fig:second}
\end{subfigure}
\hfill
\begin{subfigure}{\textwidth}
    \includegraphics[width=\textwidth]{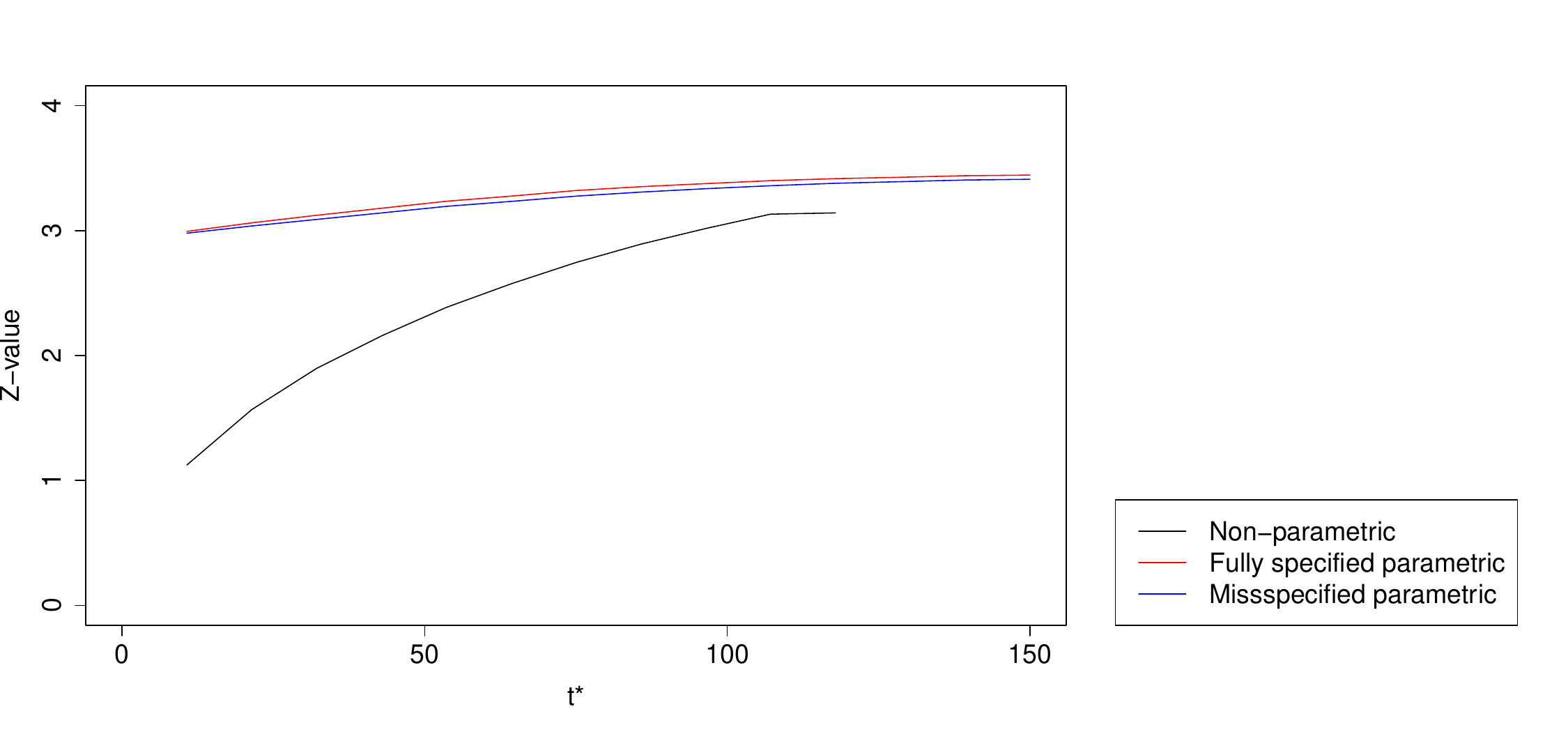}
    \caption{Mean Z-statistic for varying $\beta_3$.}
    \label{fig:third}
\end{subfigure}
\caption{Power breakdown for varying $t^*$ for each RMST estimator for $N=10^5$ simulations.}
\end{figure}

\begin{figure}[ht!]
\centering
\begin{subfigure}{0.45\textwidth}
    \includegraphics[width=\textwidth]{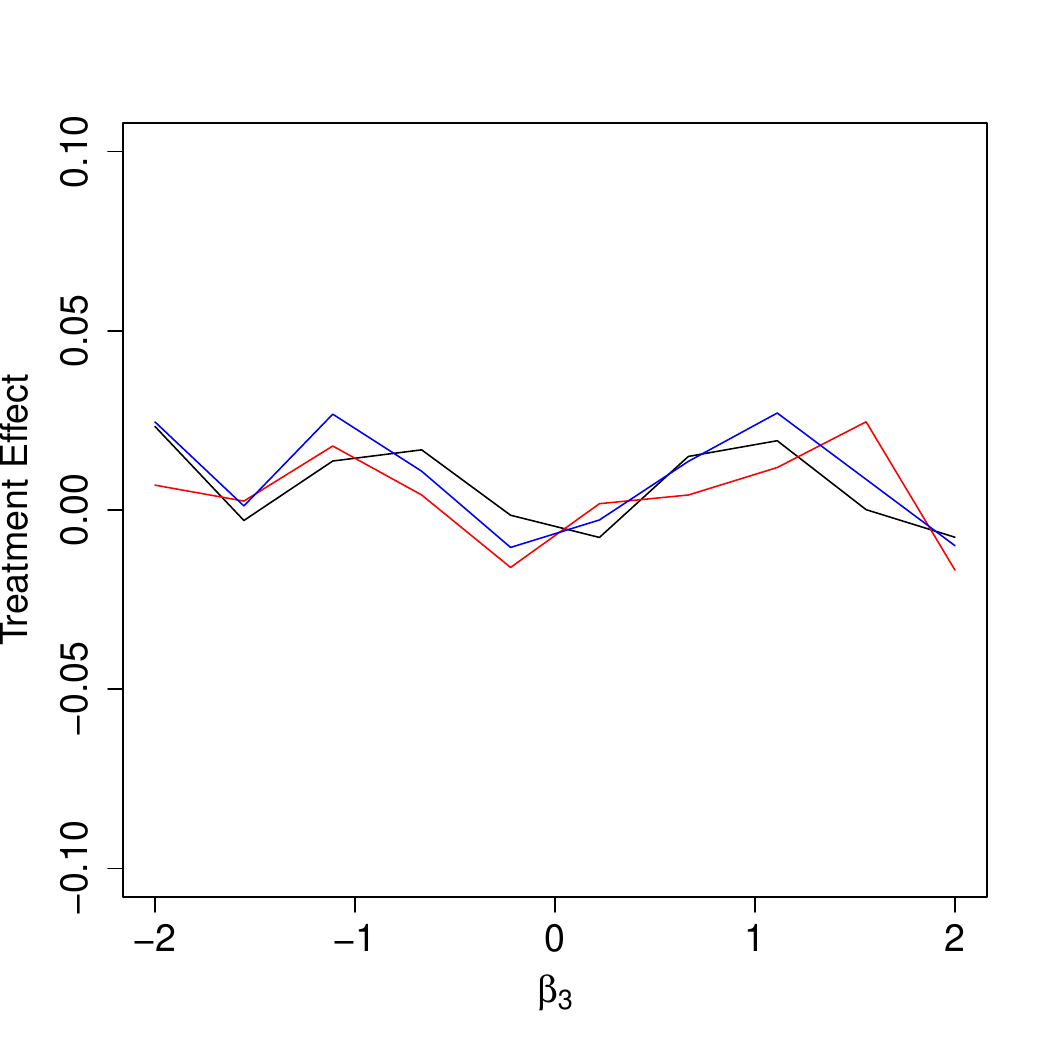}
    \caption{Mean treatment effect for varying $\beta_3$.}
    \label{fig:first}
\end{subfigure}
\hfill
\begin{subfigure}{0.45\textwidth}
    \includegraphics[width=\textwidth]{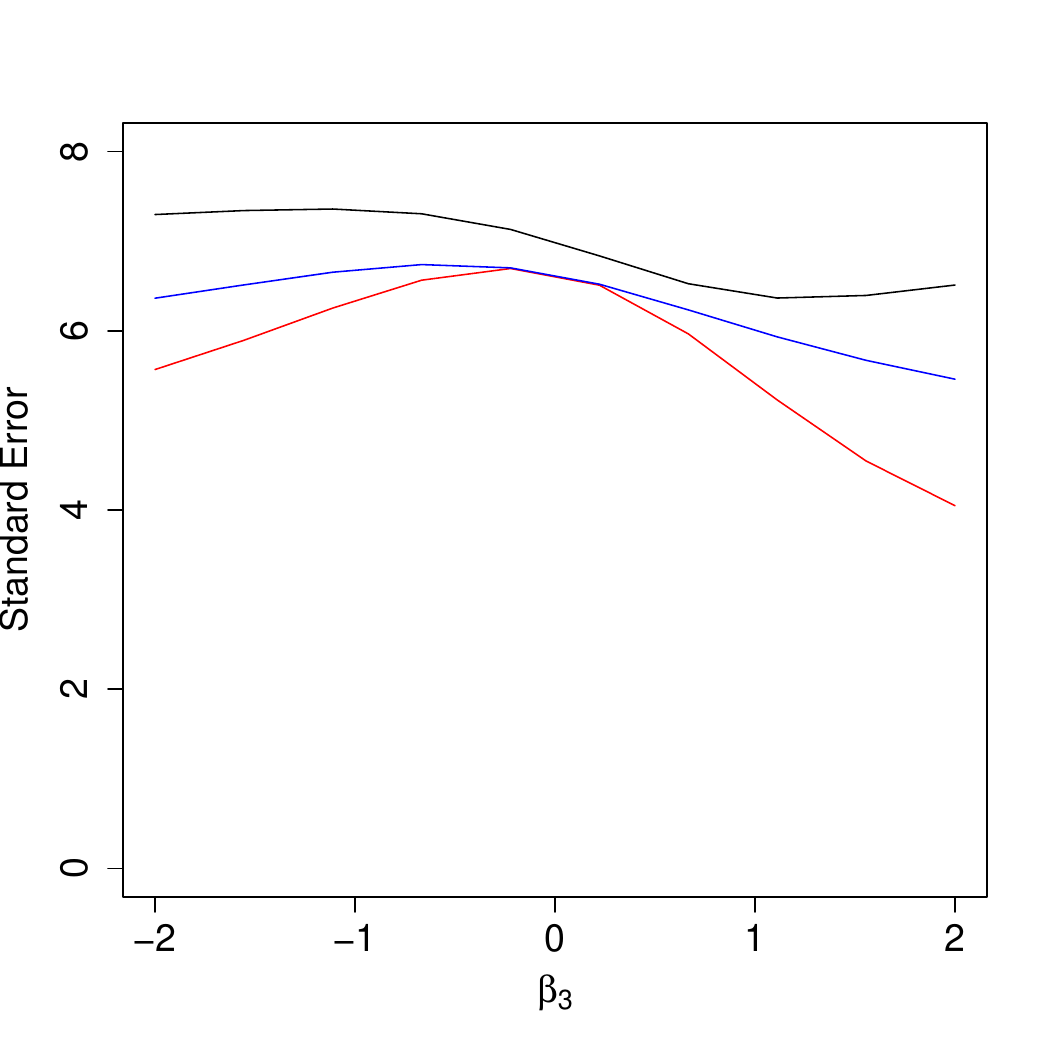}
    \caption{Mean standard error for varying $\beta_3$.}
    \label{fig:second}
\end{subfigure}
\hfill
\begin{subfigure}{\textwidth}
    \includegraphics[width=\textwidth]{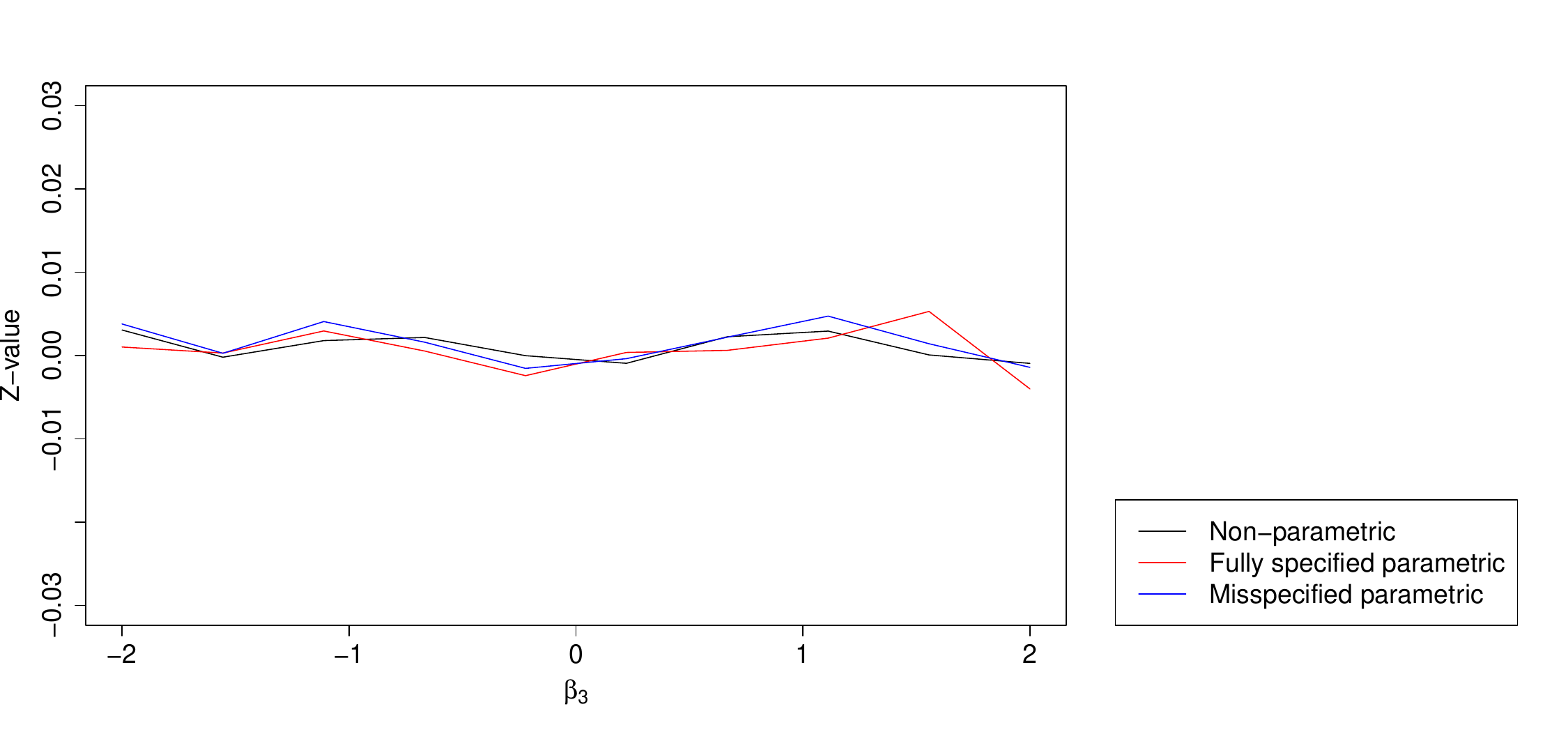}
    \caption{Mean Z-statistic for varying $\beta_3$.}
    \label{fig:third}
\end{subfigure}
\caption{Type I error for varying $\beta_3$ for each RMST estimator for $N=10^5$ simulations.}
\end{figure}

\begin{figure}[ht!]
\centering
\begin{subfigure}{0.45\textwidth}
    \includegraphics[width=\textwidth]{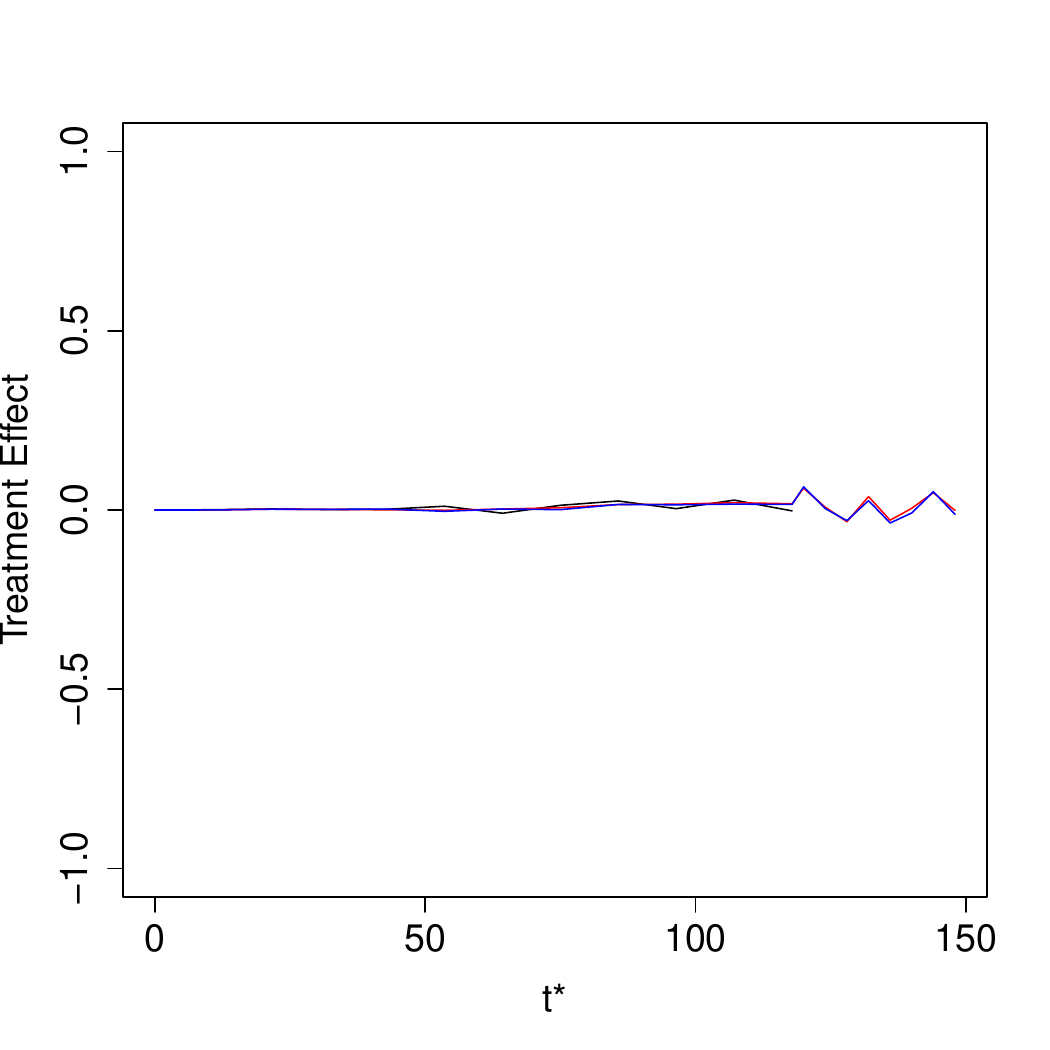} 
    \caption{Mean treatment effect for varying $t^*$.}
    \label{fig:first}
\end{subfigure}
\hfill
\begin{subfigure}{0.45\textwidth}
    \includegraphics[width=\textwidth]{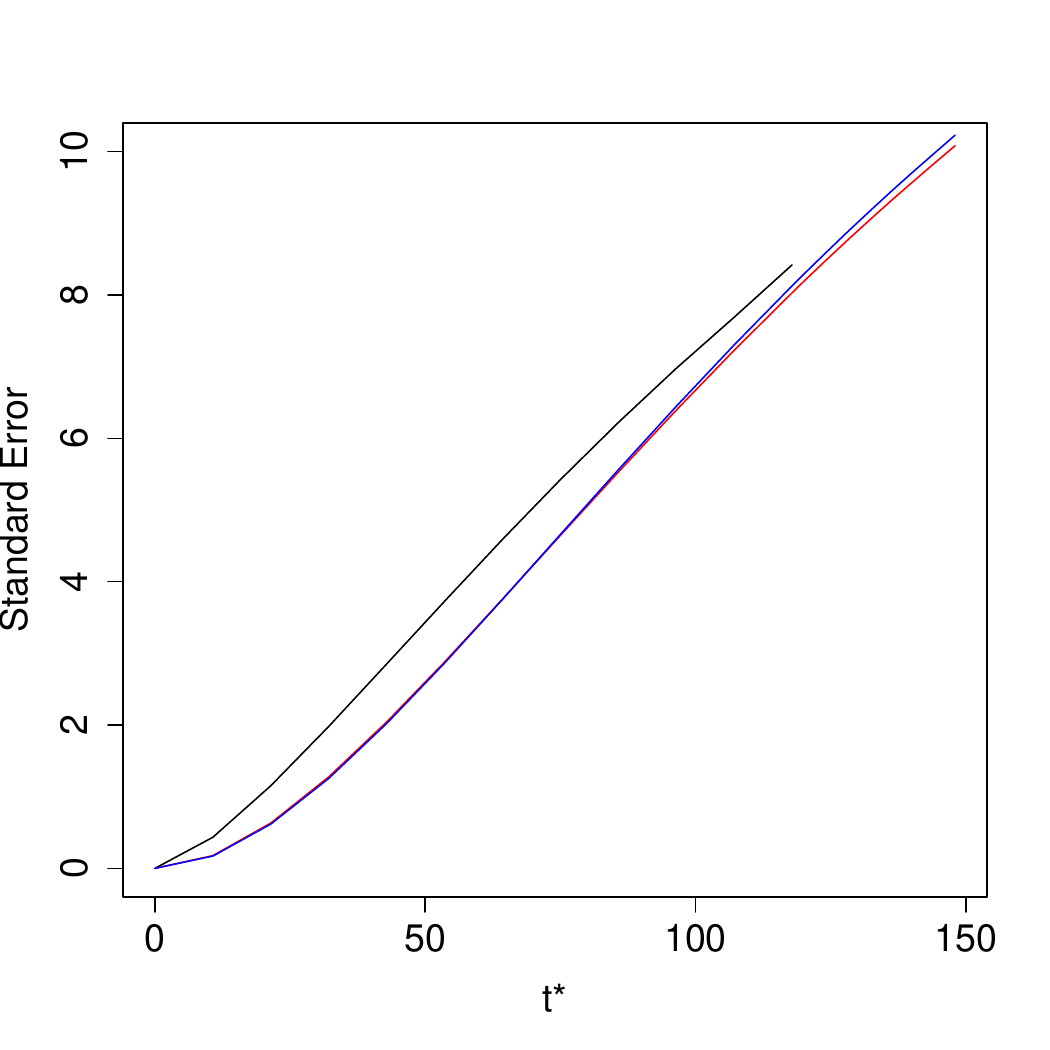}
    \caption{Mean standard error for varying $t^*$.}
    \label{fig:second}
\end{subfigure}
\hfill
\begin{subfigure}{\textwidth}
    \includegraphics[width=\textwidth]{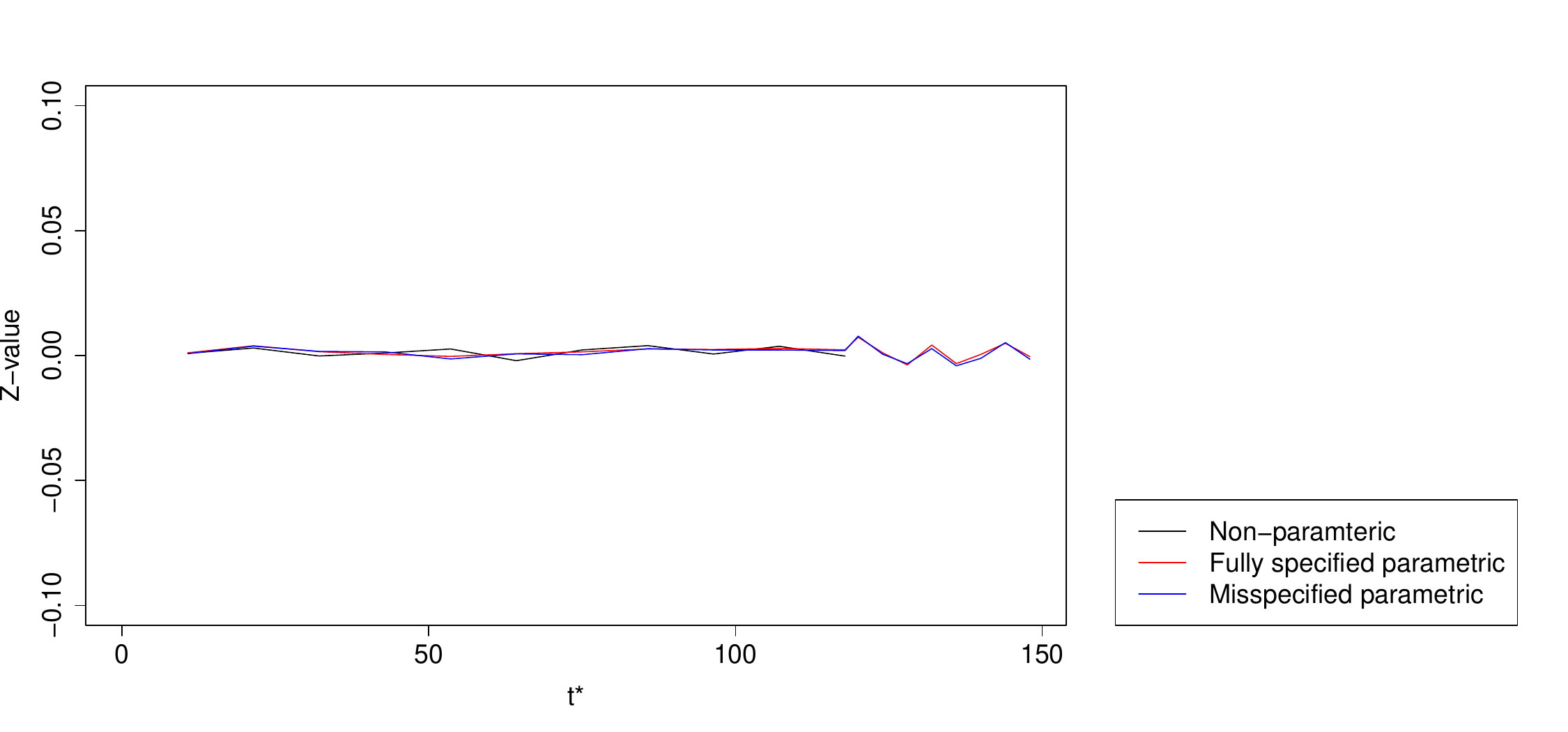}
    \caption{Mean Z-statistic for varying $t^*$.}
    \label{fig:third}
\end{subfigure}
        
\caption{Type I error for varying $t^*$ for each RMST estimator for $N=10^5$ simulations.}
\end{figure}

\clearpage

\FloatBarrier
\subsection{Crossing survival curves in the alternative direction}\label{secs5}

We investigate the effects of varying \(t^*\) when RMST is used to analyse a trial with crossing survival curves. Initially the control is considered more effective than the experimental treatment but then becomes less effective at a knot point and the survival curves cross. The curves cross in the opposite direction than in the main text and the results are obtained using the same simulation results but by permuting the treatment labels. Figure~\ref{fig:cross_KM} shows an example Kaplan-Meier plot such as this trial.

\begin{figure}[ht!]
\centering\includegraphics[width=0.9\textwidth]{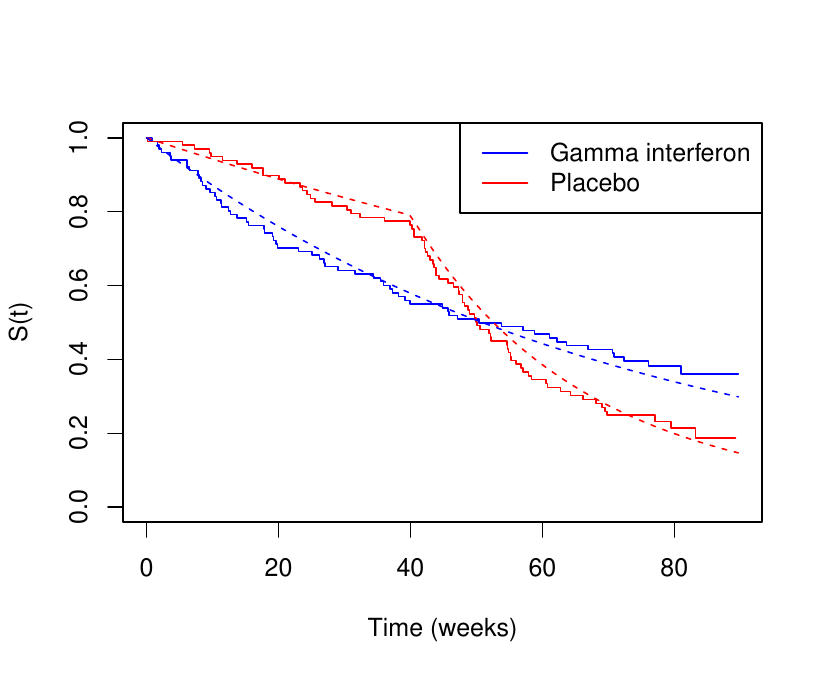}
\caption{A Kaplan-Meier plot showing the survival of 200 patients receiving either Gamma interferon or control treatment for survival distribution with crossing survival curves. Dashed lines show the parametric value of the survival function.}
\label{fig:cross_KM_supp}
\end{figure}

\begin{figure*}[!htb]
     \centering
     \begin{subfigure}[t]{0.49\textwidth}
         \centering
         \includegraphics[width=\textwidth]{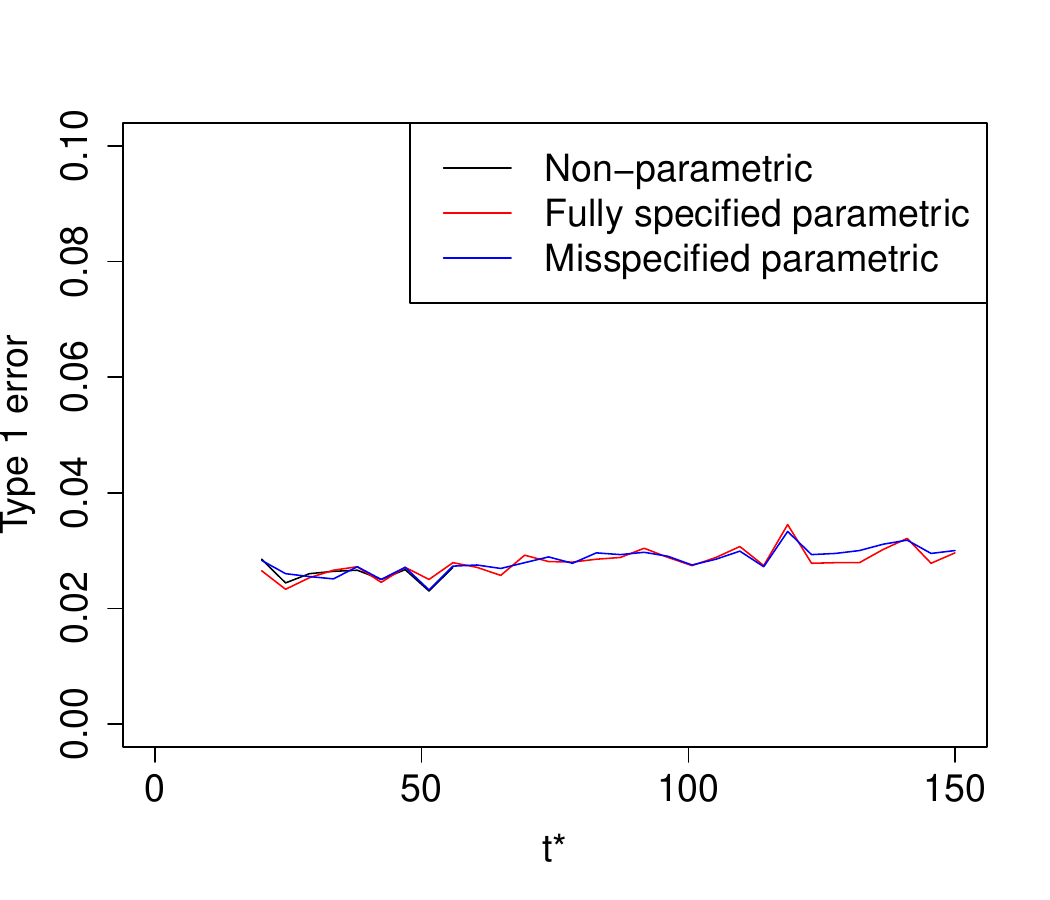}
    \end{subfigure}
     \hfill
     \begin{subfigure}[t]{0.49\textwidth}
         \centering
         \includegraphics[width=\textwidth]{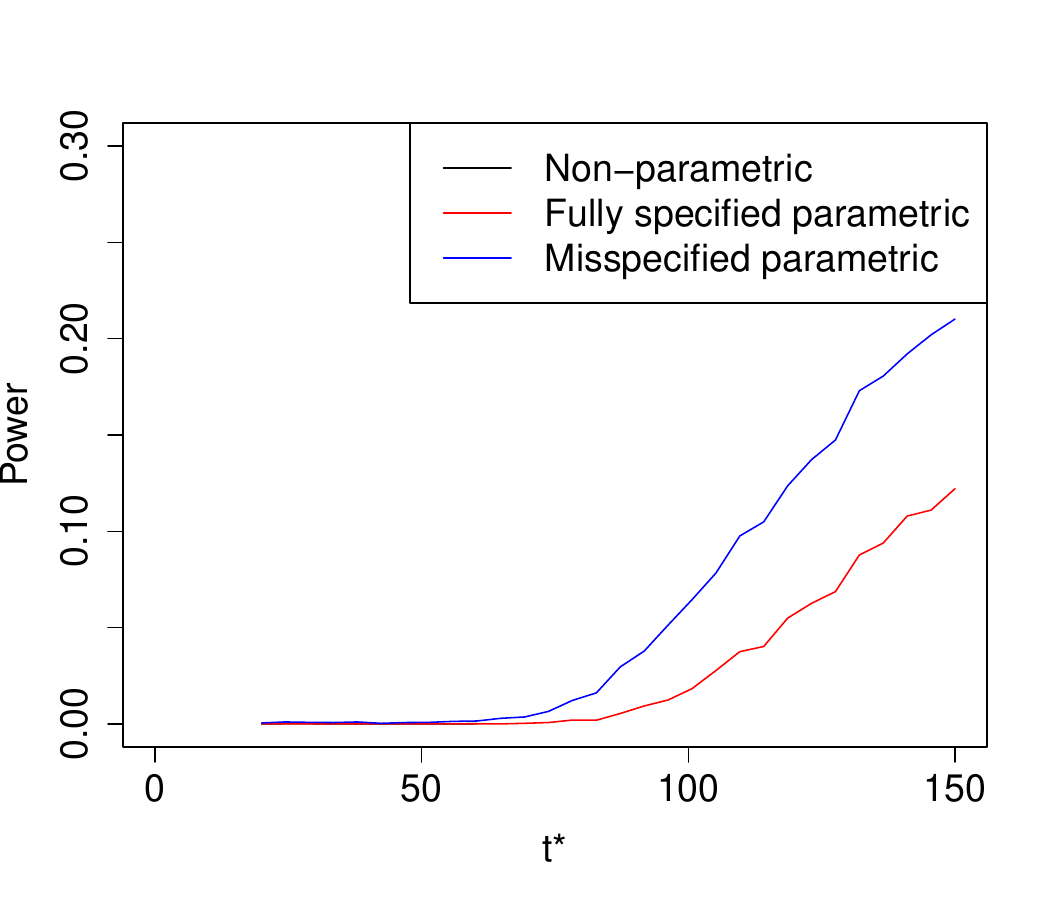}
     \end{subfigure}
        \caption{Simulation study results showing type I error rates and power when data is fit to a Cox proportional hazards model with crossing survival curves. Varying $t^*$ with fixed knot-point $t_1=40$ and misspecified knot-point $\tilde{t}_1=50$.}
        \label{fig:sim_crossing}
\end{figure*}

 Figure~\ref{fig:sim_crossing} shows the results of this simulation study. We see that the type I error rates are unaffected but the power is highly susceptible to changes in \(t^*\). As $t^*$ increases beyond the crossing point at approximately  50 weeks, the power of the trial increases as the experimental treatment becomes more effective. In general, power is low at approximately 20\% for $t^*$ at 150 weeks because the treatment is initially not working as well as placebo. We can assume that power increases further beyond 150 weeks. Under this model, the misspecified parametric endpoint appears more powerful than the correctly specified parametric endpoint. The misspecified knot point $\tilde{t}_1=50$ is later than the true knot point $t_1=40$. This means that the survival probabilities are underestimated on the placebo arm and the difference between treatment is smaller. 
 
\end{document}